\documentclass[%
 reprint,
 amsmath,amssymb,
 aps,
]{revtex4-1}
\usepackage{appendix}
\usepackage{graphicx}% Include figure files
\usepackage{dcolumn}% Align table columns on decimal point
\usepackage{bm}% bold math
\usepackage{tabularx}
\usepackage{setspace}
\usepackage{xcolor}
\usepackage{pdfpages}
\makeatletter
\AtBeginDocument{\let\LS@rot\@undefined}
\makeatother
\usepackage[portrait, margin=0.5in]{geometry}

\begin{document}
\renewcommand{\arraystretch}{1.5}
% \renewcommand{\baselinestretch}{2}
%\preprint{APS/123-QED}

\title{Fractional defect charges in $p$-atic liquid crystals on cones}% Force line breaks with \\

\author{Grace H. Zhang}
%\email{ghzhang@g.harvard.edu}
\affiliation{Department of Physics, Harvard University, Cambridge, MA 02138, USA.}
\author{David R. Nelson}
\affiliation{Department of Physics, Harvard University, Cambridge, MA 02138, USA.}%

\date{\today}
% \pagenumbering{gobble}
% \large 
\begin{abstract}
Conical surfaces, with a delta function of Gaussian curvature at the apex, are perhaps the simplest example of geometric frustration. We study two-dimensional liquid crystals with $p$-fold rotational symmetry ($p$-atics) on the surfaces of cones. For free boundary conditions at the base, we find both the ground state(s) and a discrete ladder of metastable states as a function of both the cone angle and the liquid crystal symmetry $p$. We find that these states are characterized by a set of fractional defect charges at the apex and that the ground states are in general frustrated due to effects of parallel transport along the azimuthal direction of the cone. 
We check our predictions for the ground state energies numerically for a set of commensurate cone angles (corresponding to a set of commensurate Gaussian curvatures concentrated at the cone apex), whose surfaces can be polygonized as a perfect triangular or square mesh, and find excellent agreement with our theoretical predictions.

\end{abstract}

\pacs{Valid PACS appear here}
\maketitle
% \tableofcontents
% \large
  
\section{Introduction \label{sec:intro}}

Liquid crystals with $p$-fold rotational symmetry (symmetry with respect to rotations by $2\pi/p$, where $p$ is a positive integer), also known as $p$-atics, make up a wide range of physical systems. Two of the most well-studied $p$-atics are two-dimensional triangular crystals, which can exhibit an intermediate hexatic phase ($p=6$) between the liquid and solid phases~\cite{nelson2002defects,nelson1979dislocation,halperin1978theory}, and thermotropic liquid crystals, which most commonly exhibit nematic ($p=2$) ordering~\cite{de1993physics}. 
More recently, colloidal experiments have been able to access other $p$-atics, including monolayers of sedimented colloidal hard spheres in the hexatic phase~\cite{thorneywork2017two}, pentatic ($p=5$) and triatic ($p=3$) colloidal platelets \cite{wang2018brownian,zhao2012local,zhao2009frustrated}, and tetratic ($p=4$) suspensions of colloidal cubes and square platelets~\cite{loffler2018phase,zhao2011entropic}. 

Studies of $p$-atics have also become increasingly relevant in the context of active and biological systems. Active nematic order~\cite{marchetti2013hydrodynamics,simha2002hydrodynamic,doostmohammadi2018active} is exhibited by two-dimensional suspensions of cytoskeletal filaments and motor proteins~\cite{sanchez2012spontaneous,kumar2018tunable} and epithelial monolayers~\cite{saw2017topological,kawaguchi2017topological,blanch2018turbulent}. Most recently, four-fold orientationally ordered living tissue was discovered in the crustacean \textit{Parhyale hawaiensis}\cite{cislo2021active}. 
Computational models of epithelia have also been shown to exhibit a hexatic phase, where biological cells are orientationally ordered and yet able to flow~\cite{li2018role}. (Continuous hexatic-to-crystal transitions as found in Ref.~\cite{li2018role}, and also in equilibrium simulations of 2d Lennard-Jones particles~\cite{li2020attraction}, are especially interesting because they are accompanied by a continuously diverging 2d shear viscosity $\eta_{2d} \sim \xi^2_T$, where $\xi_T$ is the translational correlation length~\cite{nelson1979dislocation}.)

In this work, we focus on the behavior of $p$-atics on the surfaces of cones. Orientational order on curved surfaces have been studied both experimentally, from liquid crystals on shells~\cite{fernandez2007novel,lopez2011frustrated} to films of microtubules and molecular motors on lipid bilayer vesicles \cite{keber2014topology}, and theoretically~\cite{lubensky1992orientational}, by considering equilibrium textures of nematic shells~\cite{vitelli2006nematic} and the ground state configurations of hexatic order on the surfaces of vesicles and torii~\cite{lubensky1992orientational,bowick2004curvature}. 
These studies have focused primarily on smoothly curved surfaces. In contrast, there have been fewer studies on the behavior of general $p$-atic liquid crystals on surfaces with curvature \textit{singularities}, such as occurs at the apex of a cone or along the seam joining two cones together to make a bicone. As shown in this paper, such concentrations of Gaussian curvature can have drastic consequences on the surrounding surfaces, even if these have zero Gaussian curvature locally and are thus nominally flat. Conical surfaces, with a delta function of Gaussian curvature at the apex, are in fact the simplest example of ``geometrical frustration,'' the incompatibility of curved surfaces with various types of order~\cite{nelson1983order}. 

We focus here on cones, which are flat everywhere except at the apex, where the Gaussian curvature positively diverges.
Conic geometries have been examined as substrates of smectic textures~\cite{mosna2012breaking} and elastic ground states of nematic solids in response to patterned disclinations~\cite{modes2010disclination,warner2020topographic,feng2021concentrated,duffy2021shape,modes2011gaussian}. They are also of interest in biological morphogenesis, where defects or anisotropic growth can facilitate the buckling of soft and initially flat plant tissues into conventional cones or hyperbolically curved anti-cone surfaces~\cite{dervaux2008morphogenesis,muller2008conical}, and where feedback between apex defects and intrinsic geometry can facilitate growth in the basal marine invertebrate \textit{Hydra}~\cite{vafa2021active}. 
Here, we seek to understand the effect of conic substrates on the textures of $p$-atics. Note that a perfectly sharp apex is not necessary to generate the physics that we describe. As will become clear below, truncated cones can exhibit very similar features. The key ingredient is not associated with an arbitrarily sharp tip but with the sloped shape of the cone flanks, which leads to the effective Gaussian curvature at its center and distinguishes the cone from a cylinder. 

\subsection{Summary of results}

We examine $p$-atic textures on cones with free (unconstrained) boundary conditions at the base, and with apex half angle $\beta$, where $\beta \rightarrow \pi/2$ in the limit of a flat disk and $\beta \rightarrow 0$ in the limit of a very narrow cylinder (see Fig.~\ref{fig:cone_schm}a). In Sec.~\ref{sec:diff_geo}, we derive in a simple pedagogical fashion the key geometrical quantities of a cone, in particular the rotation angle induced on a vector parallel transported around the apex. In Sec.~\ref{sec:MS}, we introduce a simple generalization of the Maier-Saupe model~\cite{selinger2015introduction}, which we use to numerically simulate liquid crystals on a lattice.  

For any given cone angle $\beta$, we predict theoretically in Sec.~\ref{sec:freeBC} the ground state and a ladder of quantized higher energy metastable states for different integer values of $p=1, 2, \dots, 6$. To check these results, we perform numerical energy minimizations on a computationally convenient set of ``commensurate'' cone angles such that a triangular or square lattice can be perfectly tiled on the flanks. 
We find a nearly perfect match between theory and numerics. The ground state configurations are characterized by an intriguing table of defect charges (Table~\ref{tab:free_1}) localized at the cone apex, where the charge depends on both the cone angle and the symmetry $p$ of the liquid crystal order. In this paper, we neglect for simplicity ``crystal field'' couplings between the order parameter and the curvature tensor, which can be important for $p=1$ and $p=2$~\cite{david1987critical,mbanga2012frustrated,selinger2011monte}. 

The physical picture of the ground state for cones with free boundary conditions can be described as follows. Suppose we start from a flat disk $\beta=\pi/2$, whose ground state is simply a uniform texture aligned everywhere. Since the direction is arbitrary, there is a $2\pi$ degeneracy in the order parameter orientation. As we decrease $\beta$ to make an increasingly sharper cone, we expect that defects enter the liquid crystal from the base of the cone and go to the apex, to better match the increasing Gaussian curvature at the cone tip~\cite{bowick2000interacting}. When a complete cancellation is possible, the energy of the frustration-free system vanishes. Generally, however, there is a remaining fractional defect charge at the apex due to incomplete cancellation, leading to a frustrated ground state with nonzero energy that (in the absence of extrinsic curvature effects) diverges logarithmically with system size. 

We also find above the ground state a ladder of metastable twist states, induced by the sloped periodic boundary conditions of the cone. The physics of these metastable states resembles an XY model with twisted M\"{o}bius strip boundary conditions, reviewed in the next section. 

% using centers of concentrated Gaussian curvature to create complex topographies~\cite{feng2020evolving} (effect of different radial boundary conditions would be relevant for this). 

% For small scale structures in particular, real surface imperfections maybe be comparable to the lattice constant of the materials that are grown or sedimented onto the surfaces. 

\subsection{Metastability of an XY-model on a M\"{o}bius Strip}

To set the stage for liquid crystal textures on cones, consider a one-dimensional (1d) string of spin vectors interacting on a M\"{o}bius strip with a natural twisting frequency of $q_0$. The spins are tangent to this surface and directed along the short direction of a twisted ribbon. The Hamiltonian of this system is given in the continuum limit by,
\begin{eqnarray} \label{eq:mobius_H}
H = \frac{J}{2} \int_0^L dx \left( \frac{ d \omega(x)}{dx} - q_0 \right)^2,
\end{eqnarray}
where $J$ represents the coupling strength between neighboring spins, $q_0$ is a preferred pitch for the ribbon, $L$ is the length of the M\"{o}bius strip, and $\omega(x)$ is the scalar field indicating the angle of the vector spin order parameter at position $x$ along the strip and obeys periodic boundary conditions $\omega(x+L) = \omega(x) + 2 \pi n$, where $n = 0, \pm 1, \pm 2, \dots$. 

Upon taking a functional derivative of Eq.~\ref{eq:mobius_H}, the configurations corresponding to the local minima of the energy landscape obey $\partial_x^2 \omega = 0$, which, under the imposed periodic boundary conditions, leads to
\begin{eqnarray} \label{eq:mobi_g}
\omega(x) = \frac{2 \pi n}{L} x, \quad n = 0, \pm 1, \pm 2, \dots.
\end{eqnarray}
The configurations in Eq.~\ref{eq:mobi_g} give rise to a discrete set of energies,
\begin{eqnarray} \label{eq:mobius_E}
E^{(n)} =\frac{2 \pi^2 J}{L} \left( n - \frac{q_0 L}{2 \pi} \right)^2.
\end{eqnarray}
Thus, for a given pitch $q_0$, the ground state corresponds to the value of $n=n_0$ which minimizes the argument, $E_0 = E{(n_0)}$, 
\begin{equation} \label{eq:mobius_n0}
n_0 = \underset{n}{\mathrm{argmin}}  \left( n - \frac{q_0 L}{2 \pi} \right)^2,
\end{equation}
and the remaining configurations where $n \neq n_0$ correspond to metastable states. 
Like supercurrents in a one-dimensional superconductor~\cite{chaikin1995principles}, the metastability of these states originates from the fact that, in order to transition into the nearest lower-energy local minimum, the one-dimensional spin texture $\omega(x)$ has to either twist or untwist itself by one entire revolution, which requires that the spin magnitude goes to zero at some point. 

The metastable states that we find for textures of $p$-atics on a cone are similar in spirit to a stack of M\"{o}bius strips of variable circumferential length enclosing the apex of the cone. The cone angle, which determines the amount a vector turns under parallel transport around the cone apex, controls the analog of the natural twisting frequency $q_0$ of the M\"{o}bius strip. Unlike the 1d M\"{o}bius strip (where the energy of frustrated twists typically grow linearly with length, see Eq.~\ref{eq:mobius_E}), however, we find a topological singularity with fractional charge at the apex whose energy grows \textit{logarithmically} with the system size.

\section{Differential geometry of the cone} \label{sec:diff_geo}

In this section, we summarize the geometrical quantities for the cone essential to our calculations in the rest of the paper.

\begin{figure}[h]
    \centering
    \includegraphics[width=1\columnwidth]{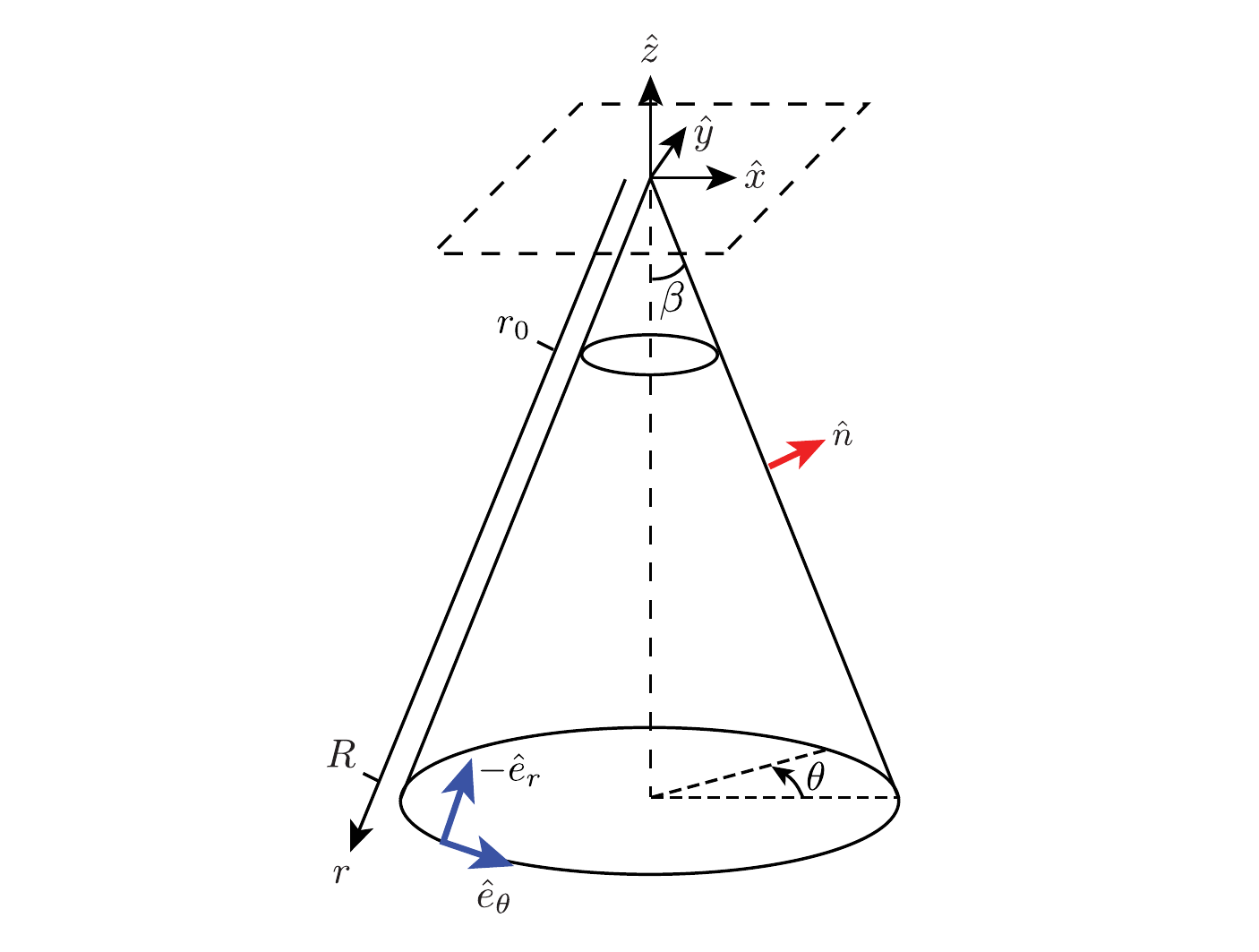}
    \caption{Schematic illustrating the notations used in this paper. Positions on the cone are parameterized by the coordinates $(r,\theta)$. Blue arrows are local surface unit tangent vectors and the red arrow is a local surface normal vector.}
    \label{fig:cone_schm}
\end{figure}

Consider a cone with apex angle $2 \beta$ with coordinates labeled in Fig.~\ref{fig:cone_schm}, where $\theta$ is the azimuthal angle around a plane perpendicular to the cone axis and $r$ is the longitudinal length along the cone surface starting from its tip, located at the origin in space.

The surface of a cone can be parameterized as (see Fig.~\ref{fig:cone_schm}),
\begin{eqnarray}
\vec R(r, \theta) = \begin{pmatrix} r \sin \beta \cos \theta \\ r \sin \beta \sin \theta \\ - r \cos \beta \end{pmatrix}.
\end{eqnarray}In the limit that the cone half-angle $\beta \rightarrow \frac{\pi}{2}$, the cone becomes a flat sheet, and in the limit $\beta \rightarrow 0 $, the cone approximates an extremely thin cylinder. 
The unnormalized tangent vectors of the conic surface are given by
\begin{eqnarray}
\partial_\theta \vec R = \begin{pmatrix} -r \sin \beta \sin \theta \\ r \sin \beta \cos \theta \\ 0 \end{pmatrix}, \quad \partial_r \vec R = \begin{pmatrix} \sin \beta \cos \theta \\ \sin \beta \sin \theta \\ - \cos \beta \end{pmatrix}
\end{eqnarray}and the corresponding metric tensor is
\begin{eqnarray} \label{eq:gab_d}
g_{\alpha \beta} = \partial_\alpha \vec R \cdot \partial_\beta \vec R = \begin{pmatrix} r^2 \sin^2 \beta & 0 \\ 0 & 1 \end{pmatrix}
\end{eqnarray}
where $x_1 = \theta, x_2 = r$, with an inverse given by,
\begin{eqnarray}
g^{\alpha \beta} = \begin{pmatrix} \frac{1}{r^2 \sin^2 \beta} & 0 \\ 0 & 1 \end{pmatrix}.
\end{eqnarray}
The outward surface normal vector is then,
\begin{eqnarray} \label{eq:norm_vec}
\hat n = -(\vec e_\theta \times \vec e_r) = \begin{pmatrix} \cos \beta \cos \theta \\ \cos \beta \sin \theta \\ \sin \beta \end{pmatrix}. 
\end{eqnarray}

To reveal and quantify the Gaussian curvature at the apex, consider a curve on the surface of the cone $\vec R(s)$, where the path parameter $s$ has units of length. 
The total curvature of this curve is given by~\cite{kamien2002geometry},
\begin{eqnarray}
\vec k &=& \frac{d \hat T(s)}{ds} = \kappa_n \hat n + \kappa_g (\hat n \times \hat T),
\end{eqnarray}
where$\hat n(s)$ is the surface normal and the unit tangent is $\hat T(s) = \frac{\vec R'(s)}{|\vec R'(s)|}$, with $\vec R'(s) = d \vec R(s)/ds$, and $\kappa_n$ and $\kappa_g$ are the normal and geodesic curvatures, respectively. 
We can obtain the normal and geodesic curvatures by taking the appropriate dot product with the total curvature vector $\vec k$: $\kappa_n =  \hat T' \cdot \hat n, \quad \kappa_g = \hat T' \cdot (\hat n \times \hat T)$. 
Upon defining $s \equiv r_0 \theta$ as our path parameter, a loop at constant longitudinal coordinate $r_0$ around the apex indicated in Fig.~\ref{fig:cone_schm} is parameterized by
\begin{eqnarray} \label{eq:Rs}
\vec R(s) = r_0 \begin{pmatrix} \sin \beta \cos \left( s/r_0 \right) \\
\sin \beta \sin \left(s/r_0 \right) \\ - \cos \beta \end{pmatrix}. 
\end{eqnarray}
The unit tangent vector $\hat T$ of the curve is then,
\begin{eqnarray}
\hat T(s) = \begin{pmatrix} - \sin \left( s/r_0 \right) \\
\cos \left(s/r_0 \right) \\ 0 \end{pmatrix},
\end{eqnarray}
which has a derivative given by
\begin{eqnarray}
\vec k = \hat T'(s) = \frac{-1}{r_0} \begin{pmatrix} \cos \left( s/r_0 \right) \\
\sin \left(s/r_0 \right) \\ 0 \end{pmatrix}. 
\end{eqnarray}
Upon using the normal vector in Eq.~\ref{eq:norm_vec}, we have
\begin{eqnarray}
\hat n \times \hat T = \begin{pmatrix} -\sin \beta \cos \left( s/r_0 \right) \\
- \sin \beta \sin \left(s/r_0 \right) \\ \cos \beta \end{pmatrix},
\end{eqnarray}
and on taking the appropriate dot products, we obtain the normal and geodesic curvatures as, 
\begin{eqnarray}
\kappa_n = - \frac{\cos \beta}{r_0}, \quad \kappa_g = \frac{\sin \beta}{r_0}. 
\end{eqnarray}
(Note that if the curve was instead a straight line along the longitudinal direction of the cone, we would fix $\theta \rightarrow \theta_0$ to be constant and take $r_0 \rightarrow r$ in Eq.~\ref{eq:Rs} to be the path parameter $s = r$.) 
Similar calculations then show that both the normal and geodesic curvature vanish on this path. 
The integrated Gaussian curvature of the surface enclosed by the curve is given by the Gauss-Bonnet theorem as~\cite{do2016differential},
\begin{eqnarray} \label{eq:Gauss_B}
\int_M K dA + \int_{\partial M} \kappa_g ds = 2 \pi \chi,
\end{eqnarray}
where $K$ is the Gaussian curvature, $\partial M$ is the curve around the apex parameterized by Eq.~\ref{eq:Rs}, $M$ is the surface of the cone enclosed by the curve, and $\chi = 1$ is the Euler characteristic of the portion of the cone surface inside the curve. Thus, the total Gaussian curvature $S$ due to the apex of the cone is given by
\begin{eqnarray} \label{eq:gauss}
S \equiv \int_M K dA =  2 \pi (1 - \sin \beta),
\end{eqnarray}
independent of $r_0$.
Although we worked with a perfect loop around the waist of the cone, we could have arrived at Eq.~\ref{eq:gauss} by considering any arbitrary curve enclosing the apex: Upon approximating the curve as a sum of infinitesimal line segments traveling along the azimuthal direction and the longitudinal direction, all contributions from the longitudinal segments to the second term in Eq.~\ref{eq:Gauss_B} vanish since the geodesic curvatures there are zero. 

We stress that the Gaussian curvature at any point on the cone away from the apex is nevertheless zero. This fact is evident from the curvature tensor $K_{\alpha \beta} = \hat n \cdot \vec R_{\alpha \beta}$, where $\vec R_{\alpha \beta} \equiv \partial_\alpha \partial_\beta \vec R$ , $\alpha, \beta = \theta, r$, which in the cone coordinates defined above, reads
\begin{eqnarray}
K^\alpha_\beta = g^{\alpha \sigma} K_{\sigma \beta} = \begin{pmatrix} \frac{\cos \beta}{\sin \beta~ r} & 0 \\ 0 & 0 \end{pmatrix}. 
\end{eqnarray}
Thus, the Gaussian curvature $ K = \text{det}(K^\alpha_{ \beta}) =0$ vanishes and the radii of curvature $\bar R_i$ are given by
\begin{eqnarray} \label{eq:radii_curve}
\bar R_\theta = \tan (\beta) ~ r, \quad \bar R_r = \infty. 
\end{eqnarray}
Note that in the flat sheet limit $(\beta \rightarrow \pi/2$), $\bar R_\theta \rightarrow \infty$, and in the cylindrical limit, $\bar R_\theta$ is the cylinder radius. 
Although the Gaussian curvature is only nonzero at the apex, we'll see that its effects on liquid crystal ground states persist all the way down the flanks of the cone, as exemplified by the parallel transport equations in the next section. 

\subsection{Parallel transport \label{sec:trans}}

In this section, we solve the parallel transport equations for the orientation field $\hat m$ (e.g. the orientation of a dense liquid of $p$-fold symmetric molecules) along the $\hat \theta$ (latitudinal) and $\hat r$ (longitudinal) directions of the cone surface. 
These differential equations allow us to extract the rotation angle experienced by an orientational vector $\hat m$ attached to these molecules upon parallel transporting from the local frame at some initial position to that of another. Henceforth, we will set $\gamma \equiv \sin \beta$. 

\begin{figure}[htb]
    \centering
    \includegraphics[width = 1\columnwidth]{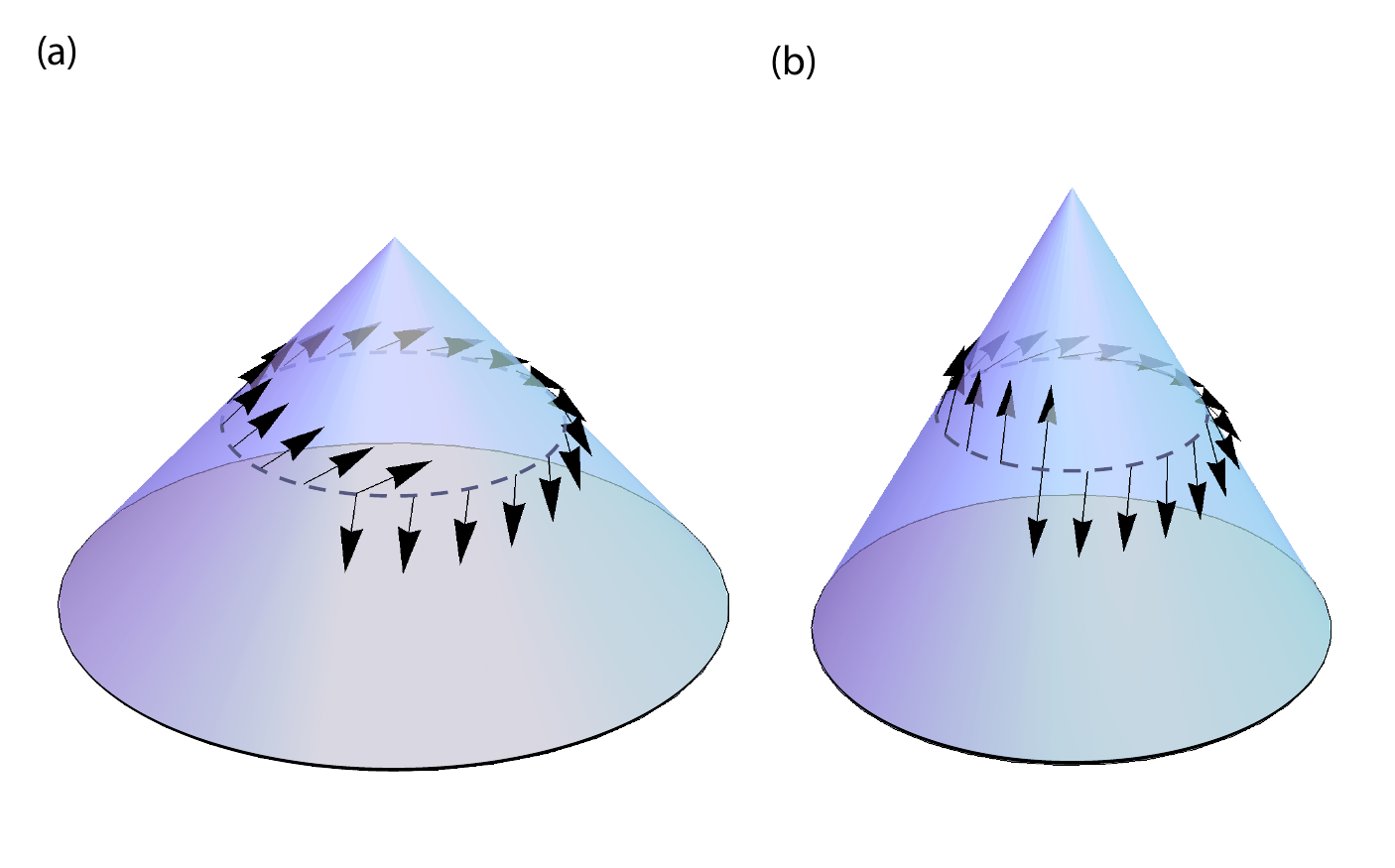}
    \caption{Parallel transport of a vector $\hat m$ on a circuit around the apex on the surface of a cone with cone angle $\gamma = \sin \beta = 2/3$ (a) and $\gamma = \sin \beta =1/2$ (b). The vector $\hat m$ suffers rotations of $240^\circ$ and $180^\circ$ in these two cases respectively. These rotations are independent of the height and orientation of these circuits. The twist in the $p$-atic order parameters as it circumnavigates the cone apex is analogous to the twisted M\"{o}bius strip discussed in Eqs.~\ref{eq:mobius_H}-\ref{eq:mobius_n0}.}
    \label{fig:PT_3d}
\end{figure}

The parallel transport of the orientational vector $m^\beta$ embedded in a $p$-fold symmetric molecule along a curve with unit tangent $u^\beta$ is given by~\cite{nakahara2018geometry},
\begin{eqnarray}
u^\alpha D_\alpha m^\beta= u^\alpha (\partial_\alpha m^\beta + \Gamma^\beta_{\alpha \gamma} m^\gamma) &=& 0  \label{eq:par_trans}
\end{eqnarray}
where $D_\alpha$ is the covariant derivative, we use the Einstein summation convention and $\Gamma^\beta_{\alpha \gamma}$ are the connection coefficients given by,
\begin{eqnarray}
\Gamma^\mu_{\nu \lambda} = \frac{1}{2} g^{\gamma \mu} \left( \partial_\nu g_{\gamma \lambda} + \partial_\lambda g_{\nu \gamma} - \partial_\gamma g_{\nu \lambda} \right).
\end{eqnarray}
On the surface of a cone, the only non-vanishing connection coefficients are,
\begin{eqnarray} \label{eq:conn_coeffs}
\Gamma^\theta_{r \theta} = \Gamma^\theta_{\theta r} = \frac{1}{r}, \quad \Gamma^r_{\theta \theta} = - r \gamma^2.
\end{eqnarray}

As detailed in Appendix~\ref{app:geo}, the rotation angle $A^\theta \sqrt{g_{\theta \theta}} d\theta = A^\theta \gamma r d \theta$ due to parallel transport in the azimuthal direction by an amount $d\theta$, when $d\theta$ is small, is given by,
\begin{eqnarray} \label{eq:parangle_theta}
 A^\theta \gamma r d\theta = \gamma d \theta.
\end{eqnarray}
In contrast, the angle of rotation due to parallel transport along the longitudinal direction by $dr$, given by $A^r \sqrt{g_{rr}} dr = A^r d r$, is always zero,
\begin{eqnarray}\label{eq:parangle_r}
A^r dr = 0.
\end{eqnarray}
The latter relation reflects the fact that the dot product of a vector with the tangent vector of a geodesic is preserved when parallel transported along a geodesic.

The total rotation angle of an orientational vector upon parallel transporting one revolution around the apex $\Omega_A = \oint \vec A \cdot d \vec l = \int_0^{2 \pi} A^\theta \gamma r d \theta$ is thus given by,
\begin{eqnarray} \label{eq:betaA}
\Omega_A = 2 \pi \gamma. 
\end{eqnarray}
By decomposing the parallel transport trajectory into azimuthal and longitudinal segments, it is easy to show that this result holds for a smooth loop of arbitrary shape encircling the cone apex. The geometrical frustration embodied in Eq.~\ref{eq:betaA} is illustrated for a polar vector in Fig.~\ref{fig:PT_3d} for $\sin \beta = \frac{2}{3}$ and $\sin \beta = \frac{1}{2}$. 
Note that for a flat disk, where $\gamma = \sin \beta = 1$, $r$ corresponds to the radial coordinate and $\theta$ corresponds to the polar angle, so the Cartesian axes of a local frame at $(r,\theta)$ coincides with the polar coordinates $\hat r, \hat \theta$, respectively. A constant orientation vector in the disk plane, with components projected onto these local frames thus appears to rotate by $2 \pi$ back to itself after one revolution around the origin. 
However, if we subtract off this artifact of polar coordinates from Eq.~\ref{eq:betaA}, we get the rotation caused by the deviation of the cone apex from flatness, $\Omega_A^\mathrm{eff} \equiv \Omega_A(\gamma) - \Omega_A(\gamma=1) = 2 \pi (\gamma - 1)$, which is, up to a sign, the integrated Gaussian curvature in Eq.~\ref{eq:gauss}.

There is a useful alternative representation of parallel transport in terms of an angular representation of the unit vector $m^\alpha(r, \theta)$, which guarantees that its norm, $g_{\alpha \beta} m^\alpha m^\beta = 1$, is preserved by parallel transport, namely
\begin{eqnarray}
(m^\theta, m^r) = \left( \frac{ \sin \omega(r, \theta)}{\gamma r},~\cos \omega(r, \theta) \right). \label{eq:omega_def}
\end{eqnarray}
The parallel transport equations derived in Appendix \ref{app:geo} for $(m^\theta, m^r)$,
\begin{eqnarray}
\partial_\theta m^r - \gamma^2 r m^\theta &=& 0 \\
\partial_\theta m^\theta + \frac{1}{r} m^r &=& 0 \\
\partial_r m^r &=& 0 \\
\partial_r m^\theta + \frac{1}{r} m^\theta &=& 0,
\end{eqnarray}
when re-expressed in terms of the angular variable $\omega(r, \theta)$, simplify to
\begin{eqnarray} \label{eq:omega_par}
\partial_\theta \omega(r, \theta) = - \gamma, \quad \partial_r \omega(r, \theta) = 0
\end{eqnarray}
with the solution 
\begin{eqnarray} \label{eq:omega_par1}
\omega(r, \theta) = \omega_0 - \gamma \theta.
\end{eqnarray}

\subsection{Commensurate cone angles}  \label{sec:comm}

For cones with certain apex angles, one can lay down a regular triangular or square mesh on the flanks of the cone without needing to introduce artifacts such as grain boundaries in the mesh lattice. We take advantage of these commensurate cone angles in our numerical simulations, described in Sec.~\ref{sec:MS}.  

We say a cone is commensurate with a triangular lattice when a section of the lattice in flat space that is a multiple of $60^\circ$ can be cut out and the remainder wrapped around to smoothly form the cone. In other words, the cone can be cut along the longitudinal direction and rolled out into a part of a regular triangular lattice. 

\begin{figure}[htb]
    \centering
    \includegraphics[width = 1\columnwidth]{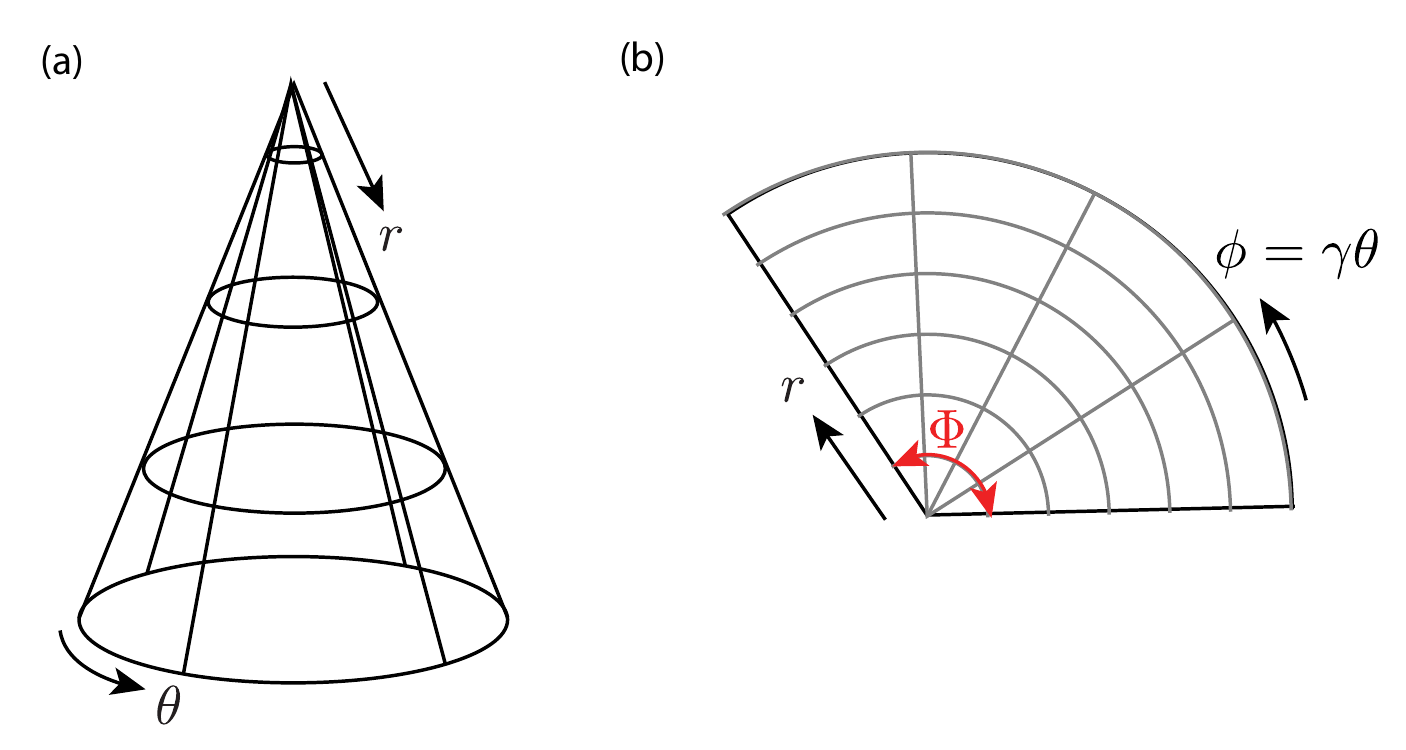}
    \caption{Coordinates $r, \theta$ of the cone (a) can be isometrically transformed onto a sector of the polar disk (b) periodic for every fixed $r$ in $\phi \rightarrow \phi + \Phi$, where $\Phi =  2 \pi \sin \beta=2 \pi \gamma$.}
    \label{fig:trans}
\end{figure}

Mathematically, we implement this rolling out procedure by a change of variables, where we can map the cone isometrically onto the polar plane (see Fig.~\ref{fig:trans}), where the local post-transformation metric is manifestly flat. 
More specifically, upon defining
\begin{eqnarray} \label{eq:trans_polar}
\quad \phi = \gamma \theta ,
\end{eqnarray}
the metric in Eq.~\ref{eq:gab_d},
\begin{eqnarray}
ds^2 = r^2 \sin^2 \beta d \theta ^2 + d r^2,
\end{eqnarray}
can be rewritten as,
\begin{eqnarray} \label{eq:metric2}
ds^2 = r^2 d\phi^2 + dr^2.
\end{eqnarray}
Eq.~\ref{eq:metric2} is the metric for a sector of a flat polar disk, with $r$ being the radial coordinate and $\phi$ being the polar angle. Note, however, that $\phi \in [0, \Phi]$, where the sector angle of the disk is given by $\Phi = 2 \pi \sin \beta\leq 2 \pi$. 

The rolled out cone, with the tip at the origin of the polar plane, maps onto a perfect slice of the triangular lattice if it corresponds to an integer number of $2 \pi/6$ radians. 
Since there are only six such angular slices available and taking all six corresponds to a flat disk, a total of five frustrated commensurate cones admit a perfectly triangulated mesh. These commensurate cone angles are given by
\begin{eqnarray}
\sin \beta^\mathrm{tri}_N = 1 - \frac{N}{6}, \quad N = 1, \dots, 5,
\end{eqnarray}
where increasing $N$ corresponds to increasing geometric frustration associated with more pointed cones. 
The parallel transport of a polar vector around the apex of two triangular lattice commensurate cones with $N=2$ and $N=3$ are shown in Fig.~\ref{fig:PT_3d}. 
A similar construction exists for square lattice meshes, where the commensurate cone angles are given by
\begin{eqnarray}
\sin \beta^\mathrm{sq}_N = 1 - \frac{N}{4}, \quad N = 1, \dots, 3. 
\end{eqnarray}
Because the cases of $N=3$ for triangular tesselations and $N=2$ for square tesselations have identical commensurate cone angles, these constructions give us a total of 7 distinct cone angles for which we can carry out numerical simulations without defects in the mesh itself.

\section{Maier-Saupe model for $p$-atic liquid crystals on a cone} \label{sec:MS}

Here, we introduce the miscroscopic model used for our numerics by generalizing the Maier-Saupe model for nematic liquid crystals on flat surfaces to that of general $p$-atics on curved surfaces. 

The original Maier-Saupe lattice Hamiltonian of a two-dimensional (2d) system of $N$ nematic liquid crystal molecules in flat space, with interactions that align nearest neighbors, is given by~\cite{selinger2015introduction},
\begin{eqnarray}
H =- J \sum_{\langle ij \rangle} \left[ \left (\hat m_i \cdot \hat m_j \right)^2 -1 \right] ,
\end{eqnarray}
where $i,j$ are site indices, $\langle ij \rangle$ indicates nearest neighbors, $\hat m_i$ is an orientational unit vector attached to a liquid crystal molecule at site $i$, and $J$ is the Maier-Saupe coupling strength between molecules at neighboring sites (note that we let $J \rightarrow  \frac{3}{2} J$, compared to the convention in Ref.~\cite{selinger2015introduction}).

Let $\omega_j$ denote the angle of the unit vector $\hat m$ in its local frame at the $j$-th site. 
For molecules with $p$-fold rotational symmetry, the interaction energy between two orientation vectors is given by the $p$-th Chebychev polynomial $T_p(x)$~\cite{zwillinger2018crc}, where $x = \hat m_i \cdot \hat m_j$ is the inner product between neighboring direction vectors,
\begin{eqnarray} 
V_{ij} = -J \left[ T_p( \hat m_i \cdot \hat m_j)  - 1 \right] = -J \left[\cos (p(\omega_i - \omega_j)) - 1 \right], \notag \\ \label{eq:V_MS_p1}
\end{eqnarray}
where $\omega_i$ is the angle of the orientation vector $\hat m_i$ in the local Cartesian frame at site $i$. 

On the surface of a cone, the vectors describing the orientation of $p$-fold symmetric molecules need to be parallel transported to the local frame of of its neighbor before their dot product is taken. The interaction energy between two neighboring molecules is hence modified by a rotation  that the molecule undergoes during the parallel transport. 

For a general $p$-atic on a regular lattice, the interaction energy for $p$-atic liquid crystals on a cone between sites $i$ and $j$ can be written,
\begin{eqnarray} \label{eq:V_MS_p2}
V_{ij} = -J \left[ \cos (p (\omega_i - \omega_j + A_{ij})) - 1 \right],
\end{eqnarray}
where $A_{ij}$ is the rotation angle of the local frame orientation vector induced by parallel transport between the $i$-th and $j$-th site. 
Equipped with the interaction energy in Eq.~\ref{eq:V_MS_p2}, we simulate $p$-atic liquid crystals on lattices on the surfaces of cones using the Python Broyden-Fletcher-Goldfarb-Shanno (BFGS) algorithm~\cite{broyden1970convergence,fletcher1970new,goldfarb1970family,shanno1970conditioning}. Our numerical energy minimizations will focus on the cone angles previously described in Sec.~\ref{sec:comm}, for which a regular polygonal mesh is especially straightforward to generate. 

Note that vectors at the cone apex do not have a well defined orientation, since the azimuthal coordinate $\theta$ is undefined there and the vector can be parallel transported by an arbitrary amount while remaining at the apex. We thus perform all energy minimizations with the orientation vector at the apex removed. 

\section{Fractional defect charge at the cone apex} \label{sec:freeBC}

In this section, we examine $p$-atic textures on conic surfaces with free boundary conditions at the cone base, showing that parallel transport on a cone leads to a vector potential term in $p$-atic energies in the continuum limit (Sec.~\ref{sec:pot}). 
As shown below in Sec.~\ref{sec:groundmeta}, the ground state configurations obtained by numerical energy minimization on cones with angles commensurate with a triangular lattice ($\sin \beta = \frac{1}{6},  \frac{2}{6},  \frac{3}{6},  \frac{4}{6},  \frac{5}{6}$) and angles commensurate with a square lattice ($\sin \beta = \frac{1}{4},  \frac{2}{4},  \frac{3}{4}$) agree with theoretical predictions in the continuum limit (Eq.~\ref{eq:qA0}). 
As tabulated in Table \ref{tab:free_1}, each ground state has either zero or a single fractional defect charge at the apex. The latter leads to a frustrated ground state with nonzero energy that grows logarithmically with the system size. In addition, we also find a ladder of metastable states, corresponding to extra ``twists'' as the $p$-atic texture wraps around the apex. (Interestingly, isolated defects do appear away from the apex when tangential boundary conditions are imposed at the cone base~\cite{vafa2022tang}.)

\subsection{XY model with a vector potential \label{sec:pot}}

On a flat 2d lattice, the energy of a $p$-fold oriented order parameter is given by~\cite{nelson2002defects,kardar2007statistical},
\begin{eqnarray} 
    H &=& - J \sum_{\langle ij \rangle} \left[ \cos (p (\omega_i - \omega_j) - 1\right]  \approx  J \sum_{\langle ij \rangle} \frac{p^2}{2} (\omega_i - \omega_j)^2, \notag \\ \label{eq:MS_coupling}
\end{eqnarray}
where $J$ is the coupling strength between two nearest-neighbor sites and the second approximate equality assumes small deformations between nearest neighbor molecules. 
In the continuum limit, the Hamiltonian becomes,
\begin{eqnarray}
H = \tilde J p^2 \int d^2 r |\vec \nabla \omega|^2,
\end{eqnarray}
where $\tilde J = J$ for a square lattice (with $4/2 = 2$ bonds per site and a lattice cell area of $a^2$) and $\tilde J = \sqrt{3} J$ for a triangular lattice (with $6/2 = 3$ bonds per site and a lattice cell area of $\sqrt{3} a^2/2$). 
Defect charges in this liquid crystal are then multiples of the minimum topological charge~\cite{nelson2002defects},
\begin{eqnarray}
\oint \vec \nabla \omega \cdot d \vec l = \frac{2 \pi s}{p}, \quad s = 0, \pm1, \dots. 
\end{eqnarray}

On a cone, the lattice Hamiltonian of a $p$-atic with nearest neighbor interactions given by Eq.~\ref{eq:V_MS_p2} is,
\begin{eqnarray} \label{eq:H_micro_cone}
H = -J \sum_{\langle i j \rangle} \left[ \cos (p (\omega_i - \omega_j + A_{ij})) - 1 \right],
\end{eqnarray}
where $\omega_i$ is the orientation angle of molecule $i$ in the local frame of site $i$ and $A_{ij}$ is the rotation angle induced by parallel transport between site $i$ and $j$. 
The continuum analog of Eq.~\ref{eq:H_micro_cone} has been discussed for general surfaces in Refs.~\cite{david1987critical} and \cite{nelson1987fluctuations},
\begin{eqnarray} \label{eq:H_vecpot}
H = \tilde J p^2 \int dr d\theta \sqrt{g} ( D_\mu m^\alpha)(D_\nu m^\beta) g^{\mu \nu} g_{\alpha \beta}.
\end{eqnarray}
where $\int dr d\theta \sqrt{g}= \int_0^{2 \pi} d\theta \int_0^R dr~r \gamma$ integrates over the surface of the cone, and $\tilde J$ depends on the microscopic structure of the lattice as described previously ($\tilde J = J$ for a square lattice and $\tilde J = \sqrt{3} J$ for a triangular lattice). We have neglected for simplicity the extrinsic curvature terms discussed in Secs.~\ref{sec:intro} and \ref{sec:conc}.

\begin{figure}[htb]
    \centering
    \includegraphics[width=1\columnwidth]{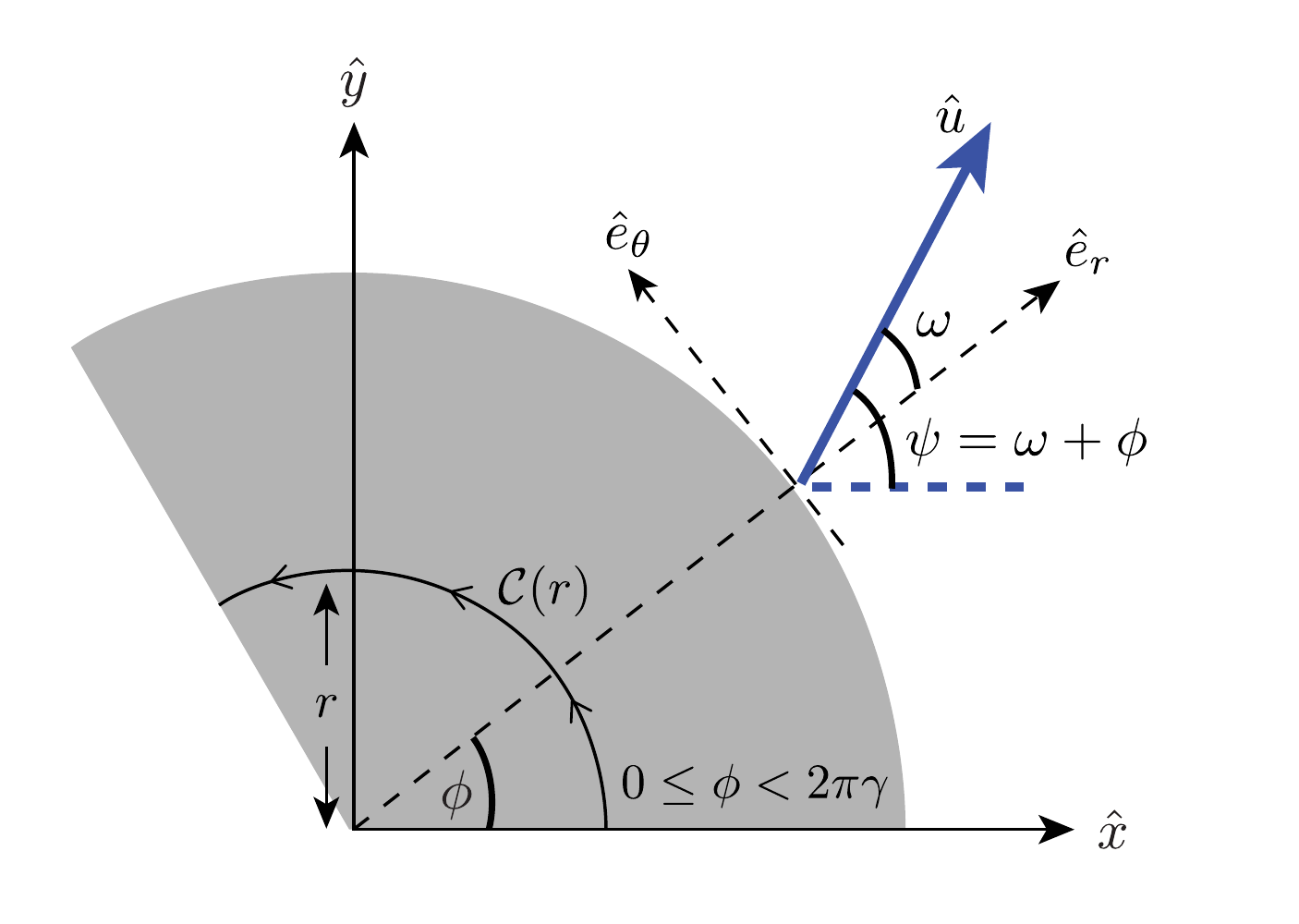}
    \caption{Relationship between the local and polar angular variables $\omega, \psi$, and the polar angle coordinate $\phi$, describing the location and orientation of the vector $\hat u$, indicated by the blue arrow, attached to our $p$-atic order parameter. The grey domain corresponds to the surface of a cone unrolled onto a flat plane. Note that $\psi = \phi + \omega$. The contour $\mathcal{C}(r)$ allows us to determine the net defect charge at the cone apex, see Eq.~\ref{eq:qA_psi}.}
    \label{fig:varphi_vartheta}
\end{figure}

As in the discussion of parallel transport in Sec.~\ref{sec:trans}, it is convenient to work in terms of an angular variable $\omega(r, \theta)$ defined by Eq.~\ref{eq:omega_def}. As illustrated in Fig.~\ref{fig:varphi_vartheta}, the angle $\omega(r,\theta)$ specifies a unit vector $\hat u = \cos [\omega (r, \theta)] \hat e_r + \sin [\omega(r, \theta)] \hat e_\theta$, where $\omega$ is the angle that $\hat u$ makes relative to the local $\hat e_r$ axis. Since $\phi = \gamma \theta$ is simply a rescaling of $\theta$, the isometric mapping of the conic surface to the polar plane identifies $\hat e_\theta$ with $\hat e_\phi$, where $\phi$ is now the polar angle of the rolled out cone, with, however, a restricted range $0 < \phi < 2 \pi \gamma$. 

It is tedious, but straightforward, to show that applying the change of variables in Eq.~\ref{eq:omega_def} to Eq.~\ref{eq:H_vecpot} leads to,
\begin{eqnarray} \label{eq:Hvartheta}
H = \tilde J p^2 \int dr d \theta \sqrt{g} \left[ \left( \frac{1}{\gamma r} \frac{ \partial \omega}{\partial \theta} + \frac{1}{r} \right)^2 + \left( \frac{ \partial \omega}{\partial r} \right)^2 \right].
\end{eqnarray}
Here, $\int dr d\theta \sqrt{g} = \int_0^R dr r \int_0^{2 \pi} d \theta \gamma$ again integrates over the surface of the cone. Note that the two squared terms in Eq.~\ref{eq:Hvartheta} are minimized when $\partial_\theta \omega = - \gamma$ and $\partial_r \omega = 0$, in agreement with the parallel transport equations in Eq.~\ref{eq:omega_par}. 

To make contact with conventional $p$-atic models in flat space, we define $\psi (r, \phi)$ as the angular field in an alternate basis, i.e. $\hat u = \cos (\psi) \hat x + \sin (\psi) \hat y$ where $\hat x$ and $\hat y$ are now fixed unit vectors pointing along the horizontal and vertical axes of the plane onto which we have unrolled our cone (see Fig.~\ref{fig:varphi_vartheta}). Simple trigonometry shows that $\psi$ is related to $\omega$ as,
\begin{eqnarray} \label{eq:varphi_vartheta}
\psi = \omega + \phi,
\end{eqnarray}
where $\phi = \gamma \theta$ is the polar angle in the plane of the unrolled cone (see Fig.~\ref{fig:trans}). 

The Hamiltonian in Eq.~\ref{eq:Hvartheta} for a cone of longitudinal length $R$ down the flanks can now be rewritten in terms of $\psi$ as
\begin{eqnarray}
H &=& \tilde J p^2 \int_0^R r dr \int_0^{2 \pi \gamma}  d\phi ~ \left|\vec \nabla \psi \right |^2,  \label{eq:Hphi}
\end{eqnarray}
where the integral is now over the gray region in Fig.~\ref{fig:varphi_vartheta} and $\vec \nabla \psi = \frac{1}{r} \frac{ \partial \psi}{\partial \phi} \hat e_\phi + \frac{ \partial \psi}{ \partial r} \hat e_r$ is the gradient in flat space polar coordinates.

\subsection{Ground and metastable twist states \label{sec:groundmeta}}

\begin{figure*}[htb]
    \centering
    \includegraphics[width=1\textwidth]{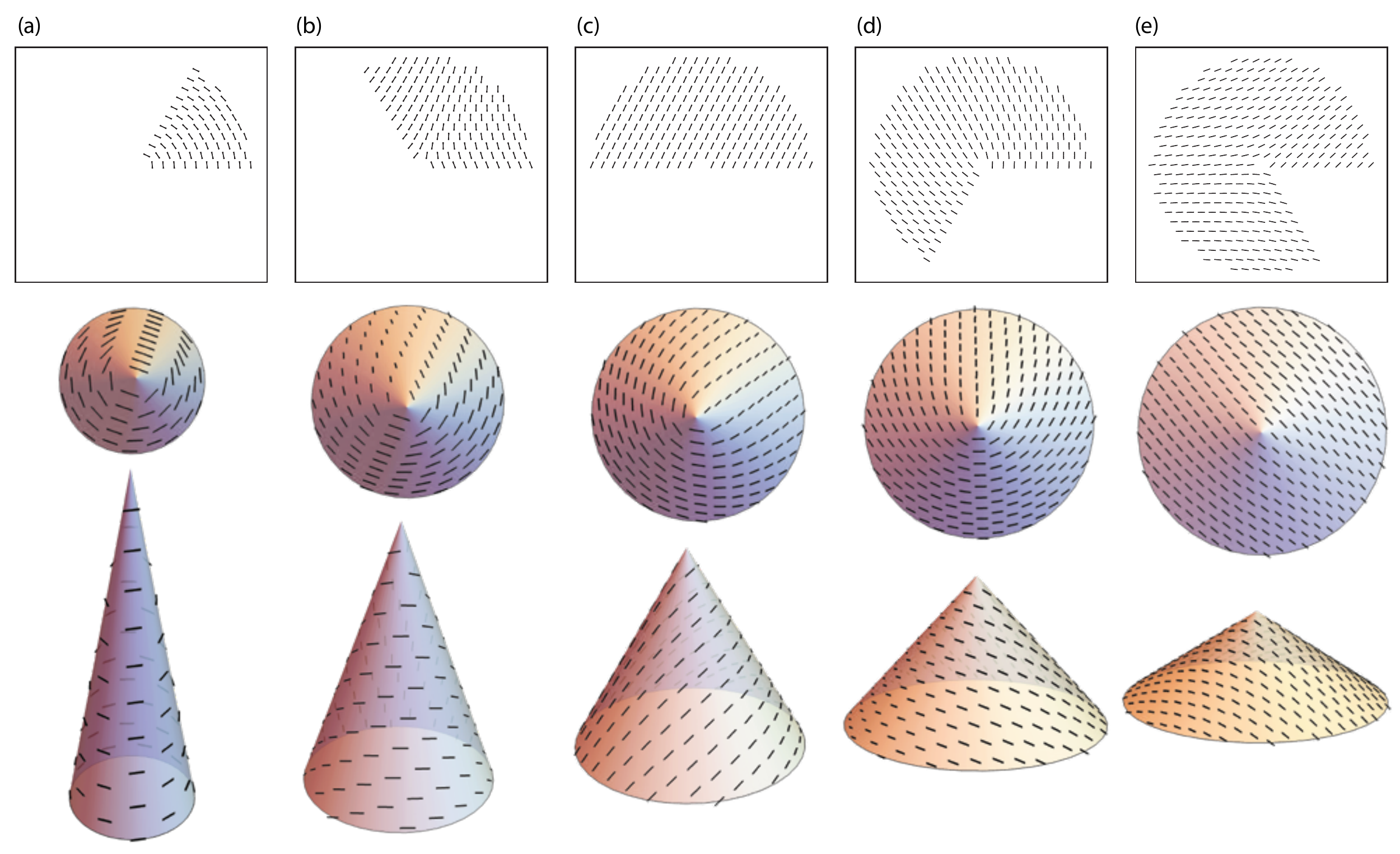}
    \caption{Ground state configurations of $p=2$ nematic liquid crystals on the surfaces of cones commensurate with a perfect triangular lattice mesh, with cone angles such that $\sin \beta=$ $1/6$ (a), $2/6$ (b), $3/6$ (c), $4/6$ (d), $5/6$ (e). The top row shows the configurations isometrically rolled onto a plane, while the middle and bottom rows show the top and side views of the texture on the surface of the cones in three dimensions.}
    \label{fig:ground}
\end{figure*}

Although Eq.~\ref{eq:Hphi} resembles a conventional 2d XY model in flat space polar coordinates $(r,\phi)$, the range of $\phi$ is restricted $0 \leq \phi \leq 2 \pi \gamma$. Hence we must pay attention to the boundary conditions on $\psi(r,\phi)$. It is easy to see from Fig.~\ref{fig:varphi_vartheta} that the boundary condition on $\omega(r,\phi)$ for $p$-atic liquid crystals reads
\begin{eqnarray} \label{eq:omega_bound}
\omega(r,\phi + 2 \pi \gamma) = \omega(r, \phi) + 2 \pi \left( \frac{s'}{p} \right), \quad s'  = 0, \pm1, \dots.
\end{eqnarray}
Note that the local twisting of the order parameter preferred by parallel transport in Eq.~\ref{eq:omega_par1} leads to $\omega(r,\phi + 2 \pi \gamma) = \omega(r,\phi) - 2 \pi \gamma$, behavior which will in general conflict with the boundary condition in Eq.~\ref{eq:omega_bound}. From Eq.~\ref{eq:varphi_vartheta}, the boundary condition on the angle $\psi(r,\phi)$ corresponding to Eq.~\ref{eq:omega_bound} is,
\begin{eqnarray} 
\psi(r,\phi+2 \pi \gamma) = \psi(r,\phi) + 2 \pi \gamma + 2 \pi \left( \frac{s'}{p} \right), \quad s' = 0,\pm1, \dots. \notag \\ \label{eq:sprime}
\end{eqnarray}
Violations of this boundary condition would lead to infinite gradient energies in Eq.~\ref{eq:Hphi}. As will be shown below (and similar to the M\"{o}bius strip problem discussed in Sec.~\ref{sec:intro}) both ground state and metastable order parameter textures for $p$-atics on the cone can be constructed by choosing a linear interpolation in $\phi$ for $\psi(r,\phi)$, consistent with the boundary condition in Eq.~\ref{eq:sprime}. The ground state corresponds to the integers that minimizes Eq.~\ref{eq:Hphi}. 

An appealing physical interpretation of these textures follows from rewriting Eq.~\ref{eq:sprime} in the following form
\begin{eqnarray}
\psi(r,\phi + 2 \pi \gamma) = \psi(r,\phi) - 2 \pi (1 - \gamma) + 2 \pi \left( \frac{s}{p} \right), ~ s = 0, \pm 1,\dots \notag \\ \label{eq:Cprime}
\end{eqnarray}
where $s = s'+p$. 
For a fixed radial coordinate $r$, the energy in Eq.~\ref{eq:Hphi} can be made zero only if $\psi(r,\phi)$ can be a constant $\psi_0$ while also obeying the boundary condition in Eq.~\ref{eq:Cprime} for $s$. This relation is equivalent to the condition that there exists an integer $s_0$ such that,
\begin{eqnarray} \label{eq:zero_energy}
\frac{s_0}{p} =1 -\gamma.
\end{eqnarray}
Recall that the total Gaussian curvature at the apex is given in Eq.~\ref{eq:gauss} as $S = 2\pi(1-\gamma)$. Therefore, we can interpret $(1 - \gamma)$ as the geometrical charge at the apex due to the background Gaussian curvature of the cone surface, while $s_0$ is the signed number of charge $+ 1/p$ defects that have entered the liquid crystal from the unconstrained cone base and moved to the apex to match the geometrical charge as much as possible. This interpretation is consistent with the fact that defects in a liquid crystal are attracted to regions of the surface whose curvature has the same sign as the defect's topological charge~\cite{turner2010vortices}. 
Note that $s_0 = 0$ for a flat disk ($\gamma = 1$), because all $p$-atic order parameters can align perfectly on a flat surface with no boundary constraints; any defects entering the liquid crystal would only increase the energy. 

The criteria in Eq.~\ref{eq:zero_energy} indeed captures all the parameter combinations of $(\gamma, p)$ for which numerical energy minimizations have produced zero ground state energies with a uniform texture $\psi = \psi_0$. 
In these special cases, the $p$-atic order parameter is parallel transported back to itself, modulo a rotation of $2 \pi / p$, upon traversing one loop around the apex. We illustrate this point in Fig.~\ref{fig:ground} for $p=2$. Here, only the commensurate cone half-angle $\beta$ such that $\sin \beta = 3/6$ is frustration free. As we show below, other values of $\sin \beta = \frac{1}{6}, \frac{2}{6}, \frac{4}{6}, \frac{5}{6}$ for $p = 2$ have frozen spin wave textures in the ground state and energies that diverge logarithmically with system size.

For combinations of $p$ and $\gamma$ where Eq.~\ref{eq:zero_energy} cannot be satisfied, the lowest energy configuration $\psi$ that obeys the appropriate periodic boundary conditions is given by,
\begin{eqnarray} \label{eq:psi_n0}
\psi(\phi) =  \left (-(1-\gamma) + \frac{s_0}{p} \right) \frac{\phi}{\gamma},
\end{eqnarray}
where $s_0$ is the value of $s$ that minimizes $|-(1-\gamma) + \frac{s}{p}|$:
\begin{equation} \label{eq:m0_def}
s_0(\gamma, p) = \underset{s}{\mathrm{argmin}}  \left|-(1-\gamma)  + \frac{s}{p}  \right| . 
\end{equation} 
If we define an effective defect charge $q_A$ at the cone apex by using the contour $\mathcal C(r)$ in Fig.~\ref{fig:varphi_vartheta}, 
\begin{eqnarray}
2 \pi q_A = \int_0^{2 \pi\gamma}\partial_\phi \psi d \phi, \label{eq:qA_psi}
\end{eqnarray}
then Eq.~\ref{eq:psi_n0} leads to a fractional value for $q_A$,
\begin{equation} \label{eq:qA0}
q_A = -(1-\gamma)  + \frac{s_0}{p}.
\end{equation}

As mentioned above, the linear dependence of $\psi(\phi)$ on $\phi$ imposes a twist on the $p$-atic order parameter for every radial coordinate $r$, similar to the M\"{o}bius strip textures embodied in Eqs.~\ref{eq:mobius_H}-\ref{eq:mobius_n0}. 
Table \ref{tab:free_1} shows the values of $q_A$ according to Eq.~\ref{eq:qA0} for all combinations of $p$ and commensurate cone angles $\gamma$ that we have examined numerically. These theoretical predictions for the defect charge at the apex match the numerical results exactly. Our results for the ground state configurations for a nematic ($p=2$) liquid crystal on cones commensurate with the triangular lattice in Fig.~\ref{fig:ground} are supplemented by plots of all numerically minimized $p$-atic textures in Appendix~\ref{app:ground}. 
Note that in some cases (e.g. $(p = 2, \gamma = 3/4)$), there can be a double degeneracy in the ground state, where there are two values of $s_0$ that both minimize the argument of Eq.~\ref{eq:m0_def} equally, leading to two values of $q_A$ with equal magnitude but opposite signs. These doublet degeneracies are also observed in our numerical ground states (see Appendix~\ref{app:ground}). 

\begin{figure*}[htb]
    \centering
    \includegraphics[width=1\textwidth]{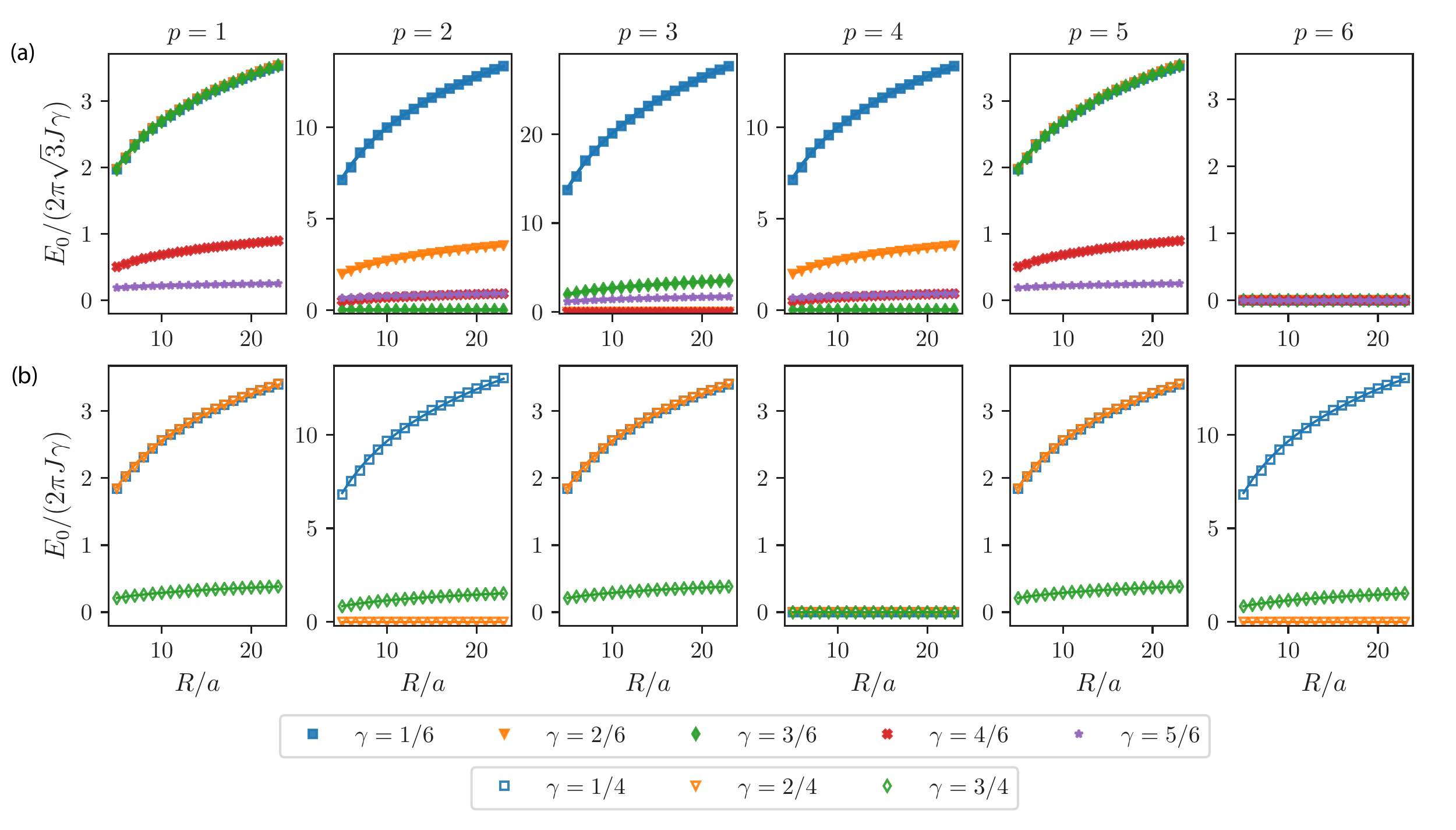}
    \caption{Ground state energies from numerical energy minimizations (colored markers) of $p$-atics as a function of the cone length $R$ at commensurate cone angles for the triangular lattice (top) and the square lattice (bottom). Colored lines are fits according to Eq.~\ref{eq:Efree_theory}, where the constant offset $E_\mathrm{core}$ is the only fitting parameter. }
    \label{fig:E_free}
\end{figure*}

\begin{table}[h!]
\centering
\begin{tabular}{c|c|c|c|c|c|c}
\hline \hline
$\sin (\beta)$ & $p = 1$ & $p = 2$ & $p=3$ & $p = 4$ & $p=5$ & $p = 6$ \\
\hline 
5/6           & $ - \frac{1}{6}$      &   $ -\frac{1}{6}$      & $ \pm \frac{ 1}{6}$ & $ \frac{1}{12}$  & $\frac{1}{30}$             & $ 0$       \\ \hline
3/4           & $-\frac{1}{4}$         & $\pm \frac{1}{4}$         & $ \frac{1}{12}$    & $0$      & $ - \frac{1}{20}$           & $\pm \frac{1}{12}$       \\ \hline
4/6          & $ - \frac{1}{2}$        & $ \frac{1}{6}$         & $ 0$              & $ -\frac{1}{12}$   & $\frac{1}{15}$               & $ 0$       \\ \hline
3/6 \text{ or } 2/4  & $ \frac{1}{2}$     & $ 0$     & $\pm \frac{1}{6}$  & $ 0$     & $ \pm \frac{1}{10}$           & $ 0$       \\ \hline
2/6           & $ \frac{1}{3}$        & $ -\frac{1}{6}$        & $0$                & $ \frac{1}{12}$    & $- \frac{1}{15}$           & $ 0$       \\ \hline
1/4          &    $\frac{1}{4}$     & $ \pm \frac{1}{4}$            & $- \frac{1}{12}$    & $0$     & $ \frac{1}{20}$           & $\pm \frac{1}{12}$      \\ \hline
1/6           & $ \frac{1}{6}$       & $ \frac{1}{6}$          & $ \pm \frac{1}{6}$ & $-\frac{1}{12}$    & $ - \frac{1}{30}$            & $ 0$     \\ \hline \hline
\end{tabular} 
\caption{Ground state apex defect charges $q_\text{A}$ according to Eq.~\ref{eq:qA0} for cone angles commensurate with the triangular and square lattices, in agreement with the numerics. \label{tab:free_1} }
\end{table}

Upon inserting Eq.~\ref{eq:psi_n0} into the Hamiltonian in Eq.~\ref{eq:Hphi}, the ground state energy of a $p$-atic on a cone with free boundary conditions is then given by
\begin{eqnarray} \label{eq:Efree_theory}
E_0 = E_\mathrm{core} +  \frac{2 \pi\tilde J p^2}{\gamma} q_A^2\ln(R/a),
\end{eqnarray}
where $q_A$ is the defect charge at the cone apex according to Eq.~\ref{eq:qA0}, $R$ is the dimension of the cone along the longitudinal direction, $a$ is the lattice constant of the polygonized mesh or some short lengthscale cutoff, and the core energy $E_\mathrm{core}$ describes the short distance physics close to the core of the defect. 

In Fig.~\ref{fig:E_free}, we plot the rescaled ground state energies $E_0/(2 \pi \gamma \tilde J)$ resulting from our $p$-atic energy minimizations, with effects of the underlying lattice ($\tilde J$) and the surface geometry of the cone ($\sim 2 \pi \gamma$) scaled out. We fit Eq.~\ref{eq:Efree_theory} to the numerical results, with the only tuning parameter being the nonuniversal constant offset of the defect core energy $E_\mathrm{core}$. The fits are plotted as colored lines in Fig.~\ref{fig:E_free} and show excellent agreement with the numerical results (colored markers).

\begin{figure}[htb]
    \centering
    \includegraphics[width=1\columnwidth]{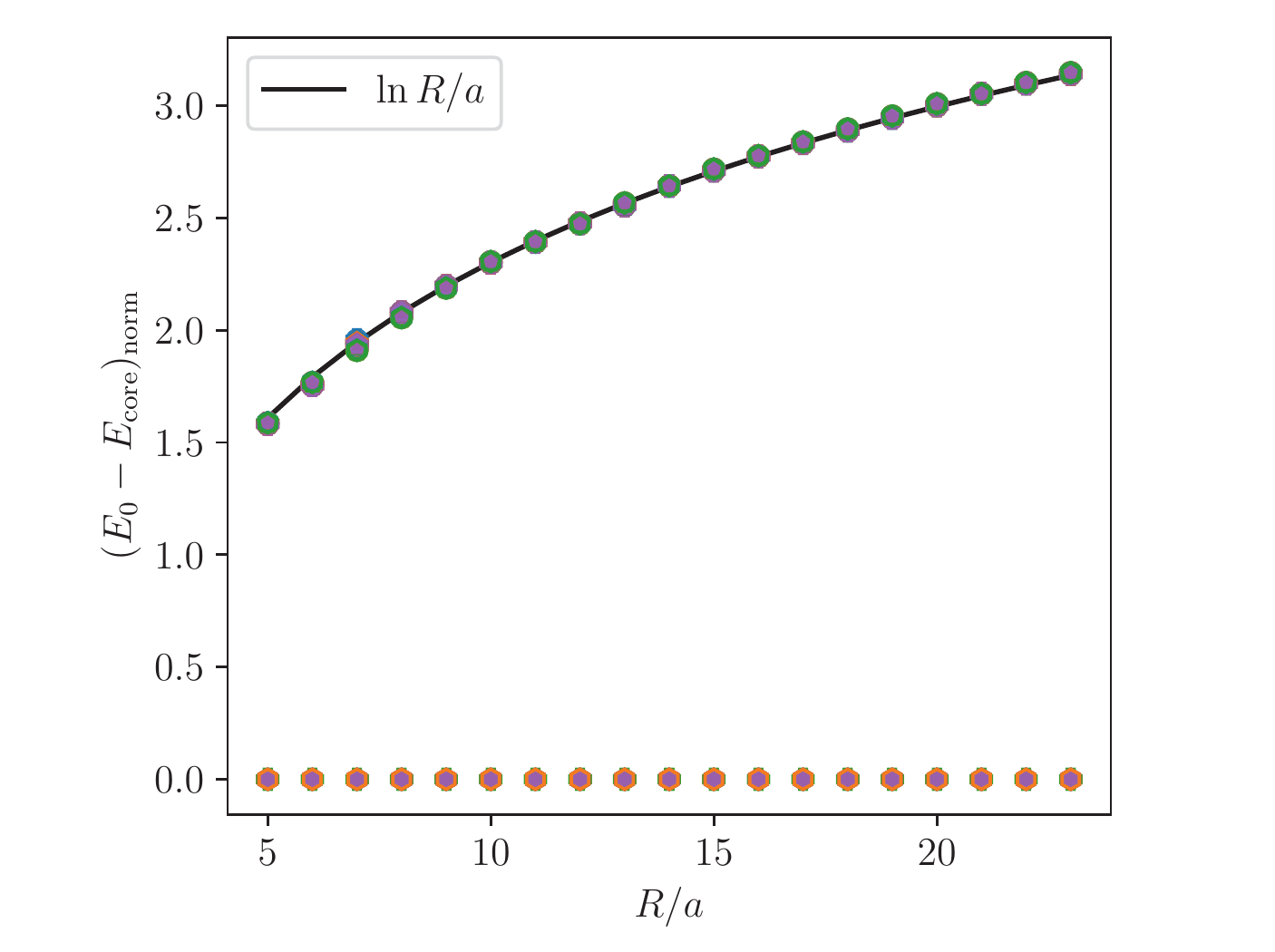}
    \caption{Ground state energies of $p$-atics for commensurate cone angles with $p=1,\dots,6$ and $\gamma = \sin \beta = \frac{1}{6}, \frac{1}{4}, \frac{2}{6}, \frac{3}{6} (= \frac{2}{4}), \frac{4}{6}, \frac{3}{4}, \frac{5}{6}$ rescaled according to Eq.~\ref{eq:Efree_theory2} collapse onto a single logarithmic curve. The only adjustable parameter is a core energy associated with the cone apex.}
    \label{fig:Efree_collapse}
\end{figure}

The numerically extracted core energies allows us to collapse the nonzero ground state energies onto a single curve,
\begin{eqnarray} \label{eq:Efree_theory2}
\frac{E_0 - E_\mathrm{core}}{2 \pi  \tilde J p^2 q_A^2/ \gamma} = \ln(R/a),
\end{eqnarray}
where $q_A$ depends on $p$ and $\gamma$ according to Eq.~\ref{eq:qA0}. The normalized energies for all parameters examined are plotted in Fig.~\ref{fig:Efree_collapse}. The collapse of the logarithmically diverging energies, when fractional apex charges are present, indeed agree with Eq.~\ref{eq:Efree_theory2}, while the zero energy cases account for situations where the cone angle is commensurate with the symmetry of the liquid crystal, i.e. those angles such that $(1-\gamma) ~\mathrm{mod}~ (1/p) = 0$. 

Recall that a $p$-atic in flat space $\gamma = 1$ with free boundary conditions would always have zero defects in its ground state. The frustration-induced fractional defect charge at the cone apex for $\gamma < 1$ is reminiscent of the appearance of a vortex line at the center of a cylinder of superfluid helium, when the superfluid goes from stationary to rotating \cite{vinen2018comparison}. The cone angle in our problem is analogous to the rotation frequency of the superfluid. 

\begin{figure*}
    \centering
    \includegraphics[width=1\textwidth]{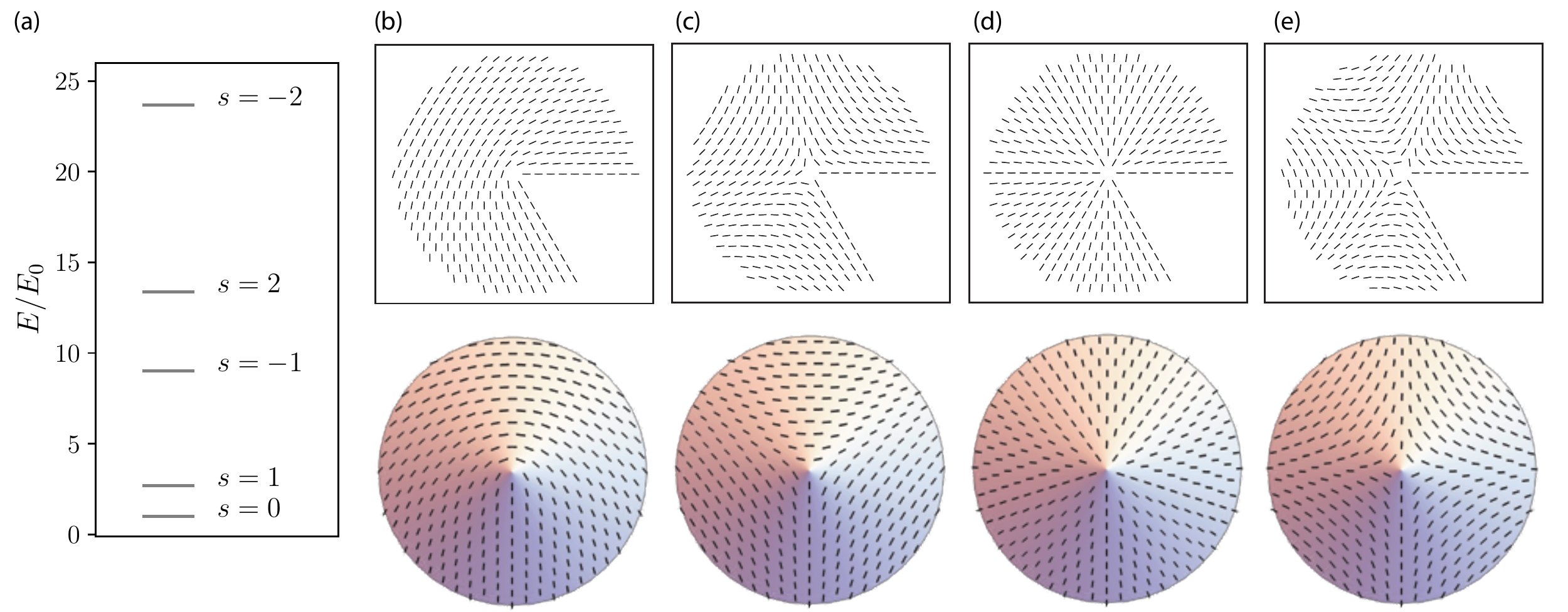}
    \caption{(a): Energies of the ground state texture and the four lowest energy metastable states for a nematic $(p=2)$ liquid crystal on a cone with angle $\sin \beta = 5/6$. Each state is labeled by an integer value of $s$ (see Eq.~\ref{eq:qA0}). Textures of the metastable states are shown in (b) $s=1$, (c) $s =-1$, (d) $s =2$, (e) $s=-2$. The top row shows the textures isometrically projected onto the plane and the bottom row shows the top view of these configurations on the surface of the cone in three dimensions. (The ground state texture with $s=0$ is shown in Fig.~\ref{fig:ground}e.)}
    \label{fig:meta}
\end{figure*}

Similar to the one-dimensional spin textures on a M\"{o}bius strip discussed in Eqs.~\ref{eq:mobius_H}-\ref{eq:mobius_n0}, the defect configurations corresponding to integers $s \neq s_0$ in Eq.~\ref{eq:qA0} result in higher energy metastable $p$-atic textures on the cone. Except for the ``central charge'' at the cone apex, these textures are all defect free on the cone flanks. For example, the four lowest energy metastable states are plotted in Fig.~\ref{fig:meta} for a nematic $(p=2)$ liquid crystal on a commensurate cone with $\gamma = 5/6$. We confirm the metastability of these states numerically by initializing the energy minimizations with the $p$-atic texture given by
\begin{eqnarray}
\psi(r, \phi) =  \left (-(1-\gamma) + \frac{s}{p} \right) \frac{\phi}{\gamma} + \eta(r, \phi), \quad s \neq s_0,
\end{eqnarray}
where $s \neq s_0$ is a fixed integer and $\eta(r,\phi)$ is a random noise, independently drawn at every site from the uniform distribution within range $[-\pi \zeta,\pi \zeta]$, where $\zeta$ represents the magnitude of the noise. Upon minimizing the energy using the BFGS algorithm for $\zeta > 0$, we observe that the apex charge of the final configuration is the same as that of the initial configuration, which confirms that the initial configuration is indeed a metastable state at a local minimum of the energy landscape. The metastable textures with lower energy are able to withstand larger magnitudes of noise than those with higher energy. For example, the first and second metastable state for $p = 2$ and $\gamma = 5/6$, represented by $s = 1$ and $s=-1$, are robust to noise magnitudes up to $\zeta \approx 0.6$ and $\zeta \approx 0.1$, respectively. 

The metastabiliy of these higher energy states arises because, in order to transform into the next lower rung of the metastable energy levels, the $p$-atic texture $\psi$ has to either twist or untwist itself by one entire revolution around the apex via the nucleation of a defect and anti-defect pair~\cite{langer1970intrinsic,ambegaokar1980dynamics}.

\section{Conclusions and outlook \label{sec:conc}}

We studied $p$-atic liquid crystals on cones at zero temperature. We found the ground states as a function of the cone half-angle $\beta$ and the molecule symmetry $p$, characterized either as uniform textures with zero energy or frustrated textures with fractional defect charges at the cone apex and an energy that diverges logarithmically with cone size. A similar set of fractional defect charges also characterize a ladder of quantized metastable liquid crystal states on the cone, reminiscent of spin textures on a M\"{o}bius strip with some natural twist wavelength similar also to a loop of supercoiled DNA~\cite{marko1994fluctuations}. Our numerical simulations were facilitated by working with a set of commensurate Gaussian curvatures concentrated at the cone apex, which allows regular tesselations of triangular or square lattices with a generalized Maier-Saupe lattice model. 

Note that the key ingredient of the physics on the cone flanks is not the arbitrarily sharp tip at which the Gaussian curvature becomes singular, but rather the Gaussian curvature hidden at the tip whether it is physically present or truncated. 
Provided one imposes free boundary conditions along the truncated rim, our results should hold for cones truncated at radius $r_0$, with $r_0 \ll R$, for which a fictitious apex defect charge sits at the center of the truncated upper rim. 

An interesting topic for future investigations would be to allow Hamiltonians with additional couplings between the molecule orientation and the extrinsic curvature of the substrate~\cite{david1987critical,selinger2011monte,napoli2012surface,mbanga2012frustrated,nguyen2013nematic}. These terms, the analog of crystal fields in spin problems, involve couplings like $\mathbf{K}_{i}^{i} \cdot \mathbf{K}_{j k} m^{j} m^{k}, \mathbf{K}_{i}^{k} \cdot \mathbf{K}_{i j} m^{j} m^{k}, \mathbf{K}_{i \ell} \cdot \mathbf{K}_{j k} m^{i} m^{\ell} m^{j} m^{k}$, where $K_i^j = K_{i k}g^{k j}$. Here, $\mathbf{K}_{ij} = K_{ij} \hat n$ and $K_{ij}$ is the curvature tensor, defined by $K_{ij} = \hat n \cdot \partial_i \partial_j \vec R$, where $\hat n$ is the normal vector. The local bending of the cone surface is given by the second fundamental form,
\begin{eqnarray} \label{eq:Kij}
K_{j}^i = \begin{pmatrix} 1/\bar R_\theta & 0 \\ 0 & 1/\bar R_r \end{pmatrix} =  \begin{pmatrix} \frac{1}{r \tan \beta} & 0 \\ 0 & 0 \end{pmatrix},
\end{eqnarray}
where the radii of curvature $\bar R_\theta$ and $\bar R_r$ are also displayed in Eq.~\ref{eq:radii_curve}. These coupling terms tend to align the orientation vector $\hat m$ along one of the substrate's principal axes of curvature, and are all proportional to $\sim \cos(2 \omega)$, where $\omega$ is the local angular variable~\cite{david1987critical}. For a $p$-atic, these additional terms will obey the appropriate $p$-fold rotational symmetry only if $2/p$ is an integer. Therefore, they are relevant for polar ($p=1$) and nematic ($p=2$) liquid crystals, but not the higher values of $p > 2$ that we have examined in this paper. Also, note from Eq.~\ref{eq:Kij} that the nonvanishing principal curvature along the azimuthal direction of the cone surface decays as we move down the flank from the apex $K_\theta = 1/\bar R_\theta \sim 1/r$. We thus expect that the effects of extrinsic curvature would diminish on a cone with its inner rim truncated far from the apex, while the intermolecular Maier-Saupe coupling, like that in Eq.~\ref{eq:MS_coupling}, retains the same strength anywhere on the cone. 
For example, if a term like $\Delta H = \sum_{j} \frac{\gamma_0}{r_j} \cos (2 \omega_j)$ is added to Eq.~\ref{eq:MS_coupling}, we would expect interfaces of dimension $\sqrt{J r / \gamma_0}$ in what was originally a M\"{o}bius strip texture without the external coupling. 
A rough estimate of the total energy integrated over the entire surface of the cone gives $E \sim \sqrt{J \gamma_0 R}$ for frustrated liquid crystal textures on the cone, in contrast to the logarithmic behavior $E \sim \ln(R/a)$ we observed from the apex charge when $\gamma_0 = 0$. 
A more detailed treatment of extrinsic curvature effects for liquid crystals on cones, both at zero and finite temperatures, is left for future work.

The fractional defect charges we find in the ground state of the cone neglecting extrinsic curvature should have intriguing consequences for $p$-atic textures at finite temperatures. On a flat 2d surface, the zero temperature state of an XY model is defect free. However, above the Kosterlitz-Thouless transition~\cite{kosterlitz2017nobel}, pairs of oppositely charged defects unbind to proliferate in a neutral plasma. In the case of the cone, however, unbinding pairs of defects will find themselves in an environment with a nonzero apex charge. We expect a transition in the Debye-H\"{u}ckle screening of this charge at the bulk Kosterlitz-Thouless defect unbinding transition. Exceptions to this screening problem are cases where the $p$-atic order is commensurate with the cone angle and there is no net topological charge at the apex at zero temperature. 

The intriguing physics described in this paper could be experimentally investigated via methods related to double emulsions, which have been used to study nematic liquid crystals on shells~\cite{fernandez2007novel}. 
A more diverse set of $p$-atic systems can be studied using colloids~\cite{zhao2009frustrated,zhao2011entropic,zhao2012local,thorneywork2017two,loffler2018phase}, by confining polygonal platelets to the surface of a cone using roughness-controlled depletion attractions~\cite{zhao2007directing}. Such conic surfaces can be made by pulling tapered glass microcapillaries or milling with a focused ion beam~\cite{sun2020colloidal,tanjeem2020effect}.  

In this paper, we examined $p$-atic liquid crystals on cones with no constraints at the rim. Different boundary conditions at the rim can alter the behavior of defects in the ground state. For example, forcing the molecules at the base of the cone to align with the circular edge (i.e. tangential boundary conditions) leads to a topological constraint on the total defect charge within the conic surface~\cite{vafa2022tang}. 
One can also explore the consequences that other types of concentrated Gaussian curvatures have on their associated surfaces. For example, hyperbolic cones (such as those associated with an isolated 7-fold disclination in a triangular lattice that can buckle into the third dimension) have a point Delta function of \textit{negative} Gaussian curvature at its center, while ridges and bicones exhibit lines of concentrated positive Gaussian curvature. 

Finally, it will be interesting to examine the dynamics of these defects both near and well away from equilibrium. The tensor hydrodynamics of general $p$-atics in flat space has recently been investigated theoretically~\cite{giomi2021hydrodynamic,giomi2021long}, while $p=2$ nematic defects in the presence of activity has been thoroughly studied both in flat space and spherical surfaces~\cite{marchetti2013hydrodynamics,simha2002hydrodynamic,doostmohammadi2018active}. It would be intriguing to generalize these studies to study the non-equilibrium active dynamics of $p$-atics on conic surfaces, realizable in epithelial monolayers~\cite{saw2017topological,kawaguchi2017topological,blanch2018turbulent} and living tissue~\cite{cislo2021active} embedded in conic geometries.

\section{Conclusions and outlook \label{sec:conc}}

We studied $p$-atic liquid crystals on cones at zero temperature. We found the ground states as a function of the cone half-angle $\beta$ and the molecule symmetry $p$, characterized either as uniform textures with zero energy or frustrated textures with fractional defect charges at the cone apex and an energy that diverges logarithmically with cone size. A similar set of fractional defect charges also characterize a ladder of quantized metastable liquid crystal states on the cone, reminiscent of spin textures on a M\"{o}bius strip with some natural twist wavelength similar also to a loop of supercoiled DNA~\cite{marko1994fluctuations}. Our numerical simulations were facilitated by working with a set of commensurate Gaussian curvatures concentrated at the cone apex, which allows regular tesselations of triangular or square lattices with a generalized Maier-Saupe lattice model. 

Note that the key ingredient of the physics on the cone flanks is not the arbitrarily sharp tip at which the Gaussian curvature becomes singular, but rather the Gaussian curvature hidden at the tip whether it is physically present or truncated. 
Provided one imposes free boundary conditions along the truncated rim, our results should hold for cones truncated at radius $r_0$, with $r_0 \ll R$, for which a fictitious apex defect charge sits at the center of the truncated upper rim. 

An interesting topic for future investigations would be to allow Hamiltonians with additional couplings between the molecule orientation and the extrinsic curvature of the substrate~\cite{david1987critical,selinger2011monte,napoli2012surface,mbanga2012frustrated,nguyen2013nematic}. These terms, the analog of crystal fields in spin problems, involve couplings like $\mathbf{K}_{i}^{i} \cdot \mathbf{K}_{j k} m^{j} m^{k}, \mathbf{K}_{i}^{k} \cdot \mathbf{K}_{i j} m^{j} m^{k}, \mathbf{K}_{i \ell} \cdot \mathbf{K}_{j k} m^{i} m^{\ell} m^{j} m^{k}$, where $K_i^j = K_{i k}g^{k j}$. Here, $\mathbf{K}_{ij} = K_{ij} \hat n$ and $K_{ij}$ is the curvature tensor, defined by $K_{ij} = \hat n \cdot \partial_i \partial_j \vec R$, where $\hat n$ is the normal vector. The local bending of the cone surface is given by the second fundamental form,
\begin{eqnarray} \label{eq:Kij}
K_{j}^i = \begin{pmatrix} 1/\bar R_\theta & 0 \\ 0 & 1/\bar R_r \end{pmatrix} =  \begin{pmatrix} \frac{1}{r \tan \beta} & 0 \\ 0 & 0 \end{pmatrix},
\end{eqnarray}
where the radii of curvature $\bar R_\theta$ and $\bar R_r$ are also displayed in Eq.~\ref{eq:radii_curve}. These coupling terms tend to align the orientation vector $\hat m$ along one of the substrate's principal axes of curvature, and are all proportional to $\sim \cos(2 \omega)$, where $\omega$ is the local angular variable~\cite{david1987critical}. For a $p$-atic, these additional terms will obey the appropriate $p$-fold rotational symmetry only if $2/p$ is an integer. Therefore, they are relevant for polar ($p=1$) and nematic ($p=2$) liquid crystals, but not the higher values of $p > 2$ that we have examined in this paper. Also, note from Eq.~\ref{eq:Kij} that the nonvanishing principal curvature along the azimuthal direction of the cone surface decays as we move down the flank from the apex $K_\theta = 1/\bar R_\theta \sim 1/r$. We thus expect that the effects of extrinsic curvature would diminish on a cone with its inner rim truncated far from the apex, while the intermolecular Maier-Saupe coupling, like that in Eq.~\ref{eq:MS_coupling}, retains the same strength anywhere on the cone. 
For example, if a term like $\Delta H = \sum_{j} \frac{\gamma_0}{r_j} \cos (2 \omega_j)$ is added to Eq.~\ref{eq:MS_coupling}, we would expect interfaces of dimension $\sqrt{J r / \gamma_0}$ in what was originally a M\"{o}bius strip texture without the external coupling. 
A rough estimate of the total energy integrated over the entire surface of the cone gives $E \sim \sqrt{J \gamma_0 R}$ for frustrated liquid crystal textures on the cone, in contrast to the logarithmic behavior $E \sim \ln(R/a)$ we observed from the apex charge when $\gamma_0 = 0$. 
A more detailed treatment of extrinsic curvature effects for liquid crystals on cones, both at zero and finite temperatures, is left for future work.

The fractional defect charges we find in the ground state of the cone neglecting extrinsic curvature should have intriguing consequences for $p$-atic textures at finite temperatures. On a flat 2d surface, the zero temperature state of an XY model is defect free. However, above the Kosterlitz-Thouless transition~\cite{kosterlitz2017nobel}, pairs of oppositely charged defects unbind to proliferate in a neutral plasma. In the case of the cone, however, unbinding pairs of defects will find themselves in an environment with a nonzero apex charge. We expect a transition in the Debye-H\"{u}ckle screening of this charge at the bulk Kosterlitz-Thouless defect unbinding transition. Exceptions to this screening problem are cases where the $p$-atic order is commensurate with the cone angle and there is no net topological charge at the apex at zero temperature. 

The intriguing physics described in this paper could be experimentally investigated via methods related to double emulsions, which have been used to study nematic liquid crystals on shells~\cite{fernandez2007novel}. 
A more diverse set of $p$-atic systems can be studied using colloids~\cite{zhao2009frustrated,zhao2011entropic,zhao2012local,thorneywork2017two,loffler2018phase}, by confining polygonal platelets to the surface of a cone using roughness-controlled depletion attractions~\cite{zhao2007directing}. Such conic surfaces can be made by pulling tapered glass microcapillaries or milling with a focused ion beam~\cite{sun2020colloidal,tanjeem2020effect}.  

In this paper, we examined $p$-atic liquid crystals on cones with no constraints at the rim. Different boundary conditions at the rim can alter the behavior of defects in the ground state. For example, forcing the molecules at the base of the cone to align with the circular edge (i.e. tangential boundary conditions) leads to a topological constraint on the total defect charge within the conic surface~\cite{vafa2022tang}. 
One can also explore the consequences that other types of concentrated Gaussian curvatures have on their associated surfaces. For example, hyperbolic cones (such as those associated with an isolated 7-fold disclination in a triangular lattice that can buckle into the third dimension) have a point Delta function of \textit{negative} Gaussian curvature at its center, while ridges and bicones exhibit lines of concentrated positive Gaussian curvature. 

Finally, it will be interesting to examine the dynamics of these defects both near and well away from equilibrium. The tensor hydrodynamics of general $p$-atics in flat space has recently been investigated theoretically~\cite{giomi2021hydrodynamic,giomi2021long}, while $p=2$ nematic defects in the presence of activity has been thoroughly studied both in flat space and spherical surfaces~\cite{marchetti2013hydrodynamics,simha2002hydrodynamic,doostmohammadi2018active}. It would be intriguing to generalize these studies to study the non-equilibrium active dynamics of $p$-atics on conic surfaces, realizable in epithelial monolayers~\cite{saw2017topological,kawaguchi2017topological,blanch2018turbulent} and living tissue~\cite{cislo2021active} embedded in conic geometries.

\begin{acknowledgements}
We are grateful for helpful conversations with F. Vafa, S. Shankar, G. Grason, and J. Sun. G.H.Z. acknowledges support by the National Science Foundation Graduate Research Fellowship under Grant No. DGE1745303. This work was also supported by the NSF through the Harvard Materials Science and Engineering Center, via Grant No. DMR-2011754, as well as by Grant No. DMR-1608501.
\end{acknowledgements}

\appendix

\section{Derivations of differential geometric quantities on a cone} \label{app:geo}

Here, we provide the details for the calculation of the parallel transport equations on the surface of a cone. 

\subsubsection{Transport along $\hat \theta$}

We first consider transport along the $\hat \theta$ direction, on a curve parameterized, in the coordinates of Fig.~\ref{fig:cone_schm}, as,
\begin{eqnarray}
\vec v = (\theta, r_0),
\end{eqnarray}
where $\theta \in [\theta_0, \theta']$ is the path parameter and $r_0$ is a constant distance down the slope of the cone. 
The unit tangent of this curve is given by,
\begin{eqnarray} \label{eq:tangtheta}
\vec u = \frac{\partial_\theta \vec v}{ \sqrt{\partial_\theta v^i g_{ij} \partial_\theta v^j }} = \frac{(1,0)}{r_0 \sin \beta}
\end{eqnarray}

The parallel transport equations in Eq.~\ref{eq:par_trans} then become,
\begin{eqnarray}
D_\theta m^j = 0,
\end{eqnarray}
written explicitly in terms of the non-vanishing connection coefficients as,
\begin{eqnarray} \label{eq:par_theta}
\partial_\theta m^r + \Gamma^r_{\theta \theta} m^\theta &=& 0 \\
\partial_\theta m^\theta + \Gamma^\theta_{r \theta} m^r &=& 0. 
\end{eqnarray}
Upon inserting the connection coefficients in Eq.~\ref{eq:conn_coeffs}, we have,
\begin{eqnarray} 
\partial_\theta m^r - r \sin^2 (\beta) m^\theta &=& 0\label{eq:par_dtheta_r} \\
\partial_\theta m^\theta + \frac{1}{r} m^r &=& 0. \label{eq:par_dtheta_theta} 
\end{eqnarray}
Upon applying $\partial_\theta$ to these equations and simplifying, we have
\begin{eqnarray}
\partial_\theta^2 m^r + \sin^2 (\beta) m^r &=& 0 \\
\partial_\theta^2 m^\theta + \sin^2 (\beta) m^\theta &=& 0. 
\end{eqnarray}
The solutions are well-known,
\begin{eqnarray}
m^\theta(\theta) &=& A \cos (\gamma \delta \theta) + B \sin (\gamma \delta \theta) \\
\quad m^r (\theta) &=& C \cos (\gamma \delta \theta) + D \sin (\gamma \delta \theta),
\end{eqnarray}
where $\gamma \equiv \sin \beta$ and $\delta \theta \equiv \theta - \theta_0$. Upon using the initial conditions $\vec m (\theta = \theta_0) = (m_0^\theta, m_0^r)$ and Eqs.~\ref{eq:par_dtheta_r} and \ref{eq:par_dtheta_theta}, we have
\begin{eqnarray} \label{eq:trans_theta_sol}
m^\theta(\theta) &=& m_0^\theta \cos (\gamma \delta \theta) - \frac{m_0^r}{r \gamma} \sin (\gamma \delta \theta) \\ m^r(\theta) &=& m_0^r \cos (\gamma \delta \theta) + m_0^\theta \gamma r \sin (\gamma \delta \theta).
\end{eqnarray}
We checked that the inner product is preserved by the above parallel transport,
\begin{eqnarray} \label{eq:mag_pres_theta}
m^i(\theta) g_{ij} m^j(\theta) = m^i(\theta_0) g_{ij} m^j(\theta_0). 
\end{eqnarray}

The rotation angle due to transporting by $d\theta$ at constant longitudinal coordinate $r_0$, which we denote as $A^\theta \gamma r_0 d \theta$, is given by the difference between the angle of $\vec m$ with respect to the tangent vector $\vec u$ of the transport curve at $\theta$ and $\theta + d \theta$: \begin{widetext} 
\begin{eqnarray} 
A^\theta \gamma r_0 d\theta = \cos^{-1} \left[\frac{ m^i(\theta_0+d\theta) g_{ij} u^j(\theta_0+d\theta)}{\sqrt{m^i(\theta_0+d\theta) g_{ij}  m^j(\theta_0+d\theta) u^i g_{ij} u^j}} \right] 
- \cos^{-1} \left[\frac{ m^i(\theta_0) g_{ij} u^j(\theta_0)}{\sqrt{m^i(\theta_0) g_{ij}  m^j(\theta_0) u^i g_{ij} u^j}} \right], \notag \\
\end{eqnarray} \end{widetext}
where $m^i(\theta)$ is given by Eq.~\ref{eq:trans_theta_sol}. Upon inserting the tangent vector given by Eq.~\ref{eq:tangtheta} and using Eq.~\ref{eq:mag_pres_theta}, we have 
\begin{widetext}\begin{eqnarray}\label{eq:betatheta}
A^\theta \gamma r_0 d\theta = \cos^{-1} \left[\frac{ m^\theta(\theta_0+d\theta) r_0 \sin \beta }{\sqrt{m^i(\theta_0) g_{ij}  m^j(\theta_0) }} \right] 
- \cos^{-1} \left[\frac{ m^\theta(\theta_0) r_0 \sin \beta }{\sqrt{m^i(\theta_0) g_{ij}  m^j(\theta_0)}} \right]. \notag \\
\end{eqnarray}\end{widetext}

\subsubsection{Transport along $\hat r$}

We now consider transport along the longitudinal $\hat r$ direction, on the curve parameterized as (see Fig.~\ref{fig:cone_schm}),
\begin{eqnarray}
\vec v = (\theta_0, r),
\end{eqnarray}
where $r \in [r_0, r'] $ is the path parameter and $\theta_0$ is a constant azimuthal angle. 
The unit tangent of this curve is given by
\begin{eqnarray} \label{eq:tangr}
\vec u = \frac{\partial_r \vec v}{ \sqrt{\partial_r v^i g_{ij} \partial_r v^j} } = (0,1). 
\end{eqnarray}

The parallel transport equation then becomes,
\begin{eqnarray}
D_r m^j = 0,
\end{eqnarray}
written in terms of the non-vanishing connection coefficient as,
\begin{eqnarray} \label{eq:par_r}
 \partial_r m^\theta + \Gamma^\theta_{\theta r} m^\theta &=& 0. 
\end{eqnarray}
Upon substituting in Eq.~\ref{eq:conn_coeffs}, Eq.~\ref{eq:par_r} becomes,
\begin{eqnarray} 
 \partial_r m^\theta + \frac{1}{r} m^\theta &=& 0. \label{eq:par_d}
\end{eqnarray}
We integrate both sides of the equation and arrive at
\begin{eqnarray}
m^\theta r = F,
\end{eqnarray}
where $F$ is some constant.
Upon using the initial conditions $\vec m (\theta_0, r_0) = (m_0^\theta, m_0^r)$, we have
\begin{eqnarray} \label{eq:trans_r_sol}
m^\theta = \frac{r_0}{r} m_0^\theta, \quad m^r = m_0^r. 
\end{eqnarray}
We again checked that the inner product is preserved by this parallel transport,
\begin{eqnarray} \label{eq:mag_pres_r}
m^i(r) g_{ij}(r) m^j(r) = m^i(r_0) g_{ij}(r_0) m^j(r_0). 
\end{eqnarray}

The angle of rotation due to parallel transport by $dr$, denoted as $A^r d r$, is given by the difference between the angle of $\vec m$ with respect to the tangent vector $\vec u$ of the transport curve, which in this case is a geodesic, at $r$ and $r + d r$: \begin{widetext}
\begin{eqnarray} 
A^r dr = \cos^{-1} \left[\frac{ m^i(r + d r) g_{ij}(r + d r) u^j}{\sqrt{m^i(r + d r) g_{ij} (r + d r) m^j(r + d r) u^i g_{ij} u^j}} \right] 
- \cos^{-1} \left[\frac{ m^i(r) g_{ij}(r) u^j}{\sqrt{m^i(r) g_{ij}(r)  m^j(r) u^i g_{ij} u^j}} \right], \notag \\
\end{eqnarray} \end{widetext}
where $m^i(r)$ is given by Eq.~\ref{eq:trans_r_sol}. Upon inserting the tangent vector given by Eq.~\ref{eq:tangr} and using Eq.~\ref{eq:mag_pres_r}, we have 
\begin{widetext}\begin{eqnarray}
A^r dr = \cos^{-1} \left[\frac{ m^r(r + d r) }{\sqrt{m^i(r) g_{ij}(r)  m^j(r) }} \right] 
- \cos^{-1} \left[\frac{ m^r(r) }{\sqrt{m^i(r) g_{ij}(r)  m^j(r)}} \right]. \notag \\
\end{eqnarray} \end{widetext}
Since $m^r(r) = m^r(r')$ is constant along the $\hat r$ curve according to Eq.~\ref{eq:trans_r_sol}, we have that
\begin{eqnarray}\label{eq:betar}
A^r dr = 0,
\end{eqnarray}
as must be the case for parallel transport along $\hat e_r$, which is a geodesic. 

\section{Ground state textures} \label{app:ground}
The following pages show the ground state textures of $p$-atics on cones with free boundary conditions at the base, obtained from numerical energy minimizations of the Hamiltonian in Eq.~\ref{eq:H_micro_cone}. Like the apex defect charges in Table~\ref{tab:free_1}, the configurations are arranged by row according to $\sin \beta$, where $\beta$ is the half cone angle (see Fig.~\ref{fig:cone_schm}), and by column according to liquid crystal symmetry parameter $p$. 
Additionally, although $\sin \beta = 3/6$ is in value equivalent to $\sin \beta = 2/4$, the corresponding numerical ground states are shown separately here, where $\sin \beta = 3/6$ indicates simulations done on a triangular lattice mesh, while $\sin \beta = 2/4$ indicates those done on a square lattice mesh. 

Textures marked with a shaded blue star in the lower right hand corner indicate unfrustrated, zero energy, ground states. These correspond to the following combinations of parameters, each for which there exists an integer value of $s_0$ that satisfies Eq.~\ref{eq:zero_energy}: $(p = 2, ~\sin \beta = 3/6,~2/4), ~(p = 3, ~\sin \beta = 4/6, ~2/6),~(p=4, ~\sin \beta = 3/4, ~3/6, ~2/4, ~1/4), (p=6, ~\sin \beta = 5/6, ~4/6, ~3/6, ~2/4, ~2/6, ~1/6)$. Certain parameter sets have doublet degeneracies in the ground state. These are: $(p=2, ~\sin \beta = 3/4,~ 1/4),~(p = 3, ~\sin \beta = 5/6, ~3/6, ~2/4, ~1/6),~(p = 5, ~\sin \beta = 3/6, ~2/4),~(p=6, ~\sin \beta = 3/4, ~1/4)$. 

\begin{figure*}
    \centering
    \includegraphics[width=1\textwidth]{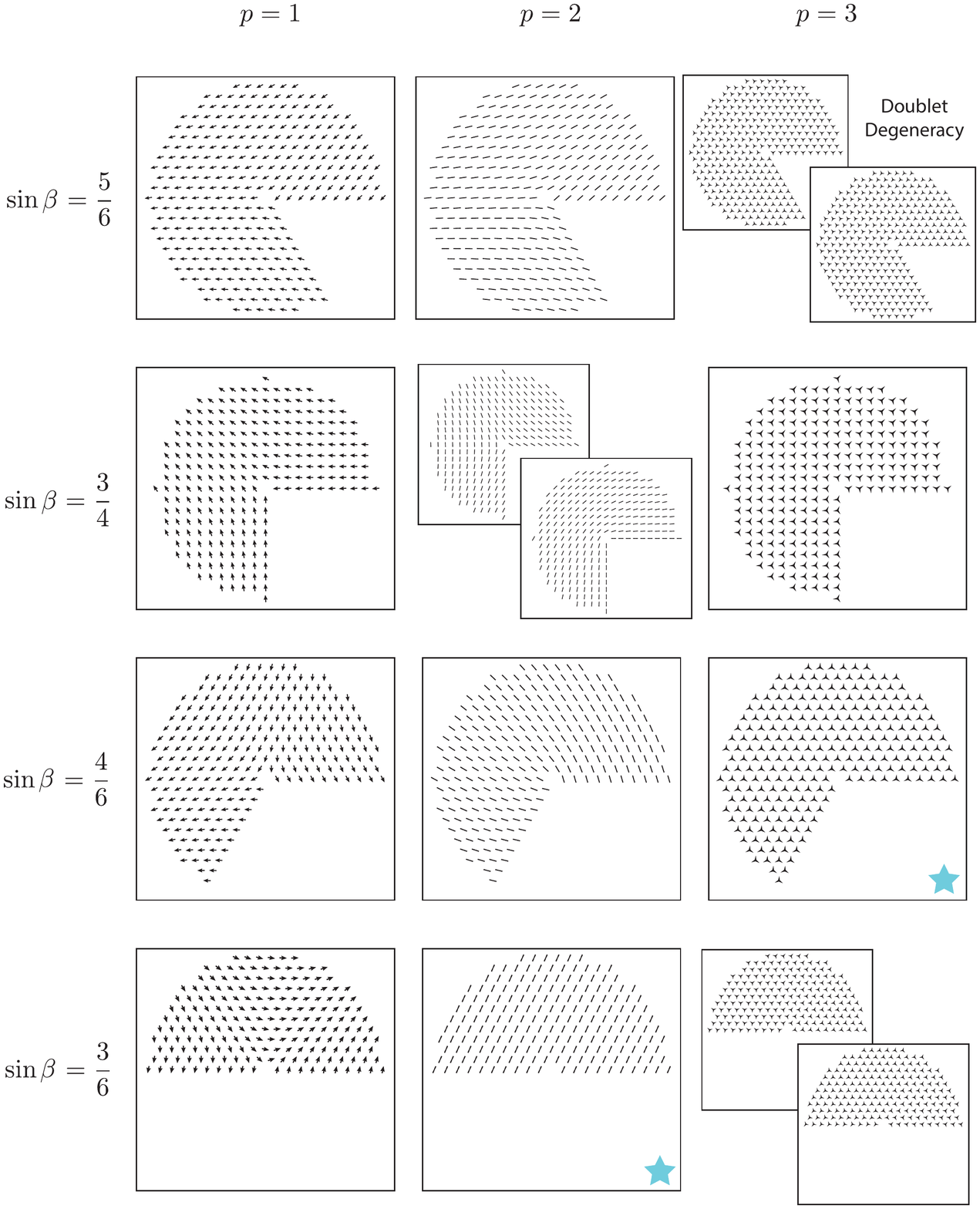}
    % \caption{Caption}
    % \label{fig:my_label}
\end{figure*}

\begin{figure*}
    \centering
    \includegraphics[width=1\textwidth]{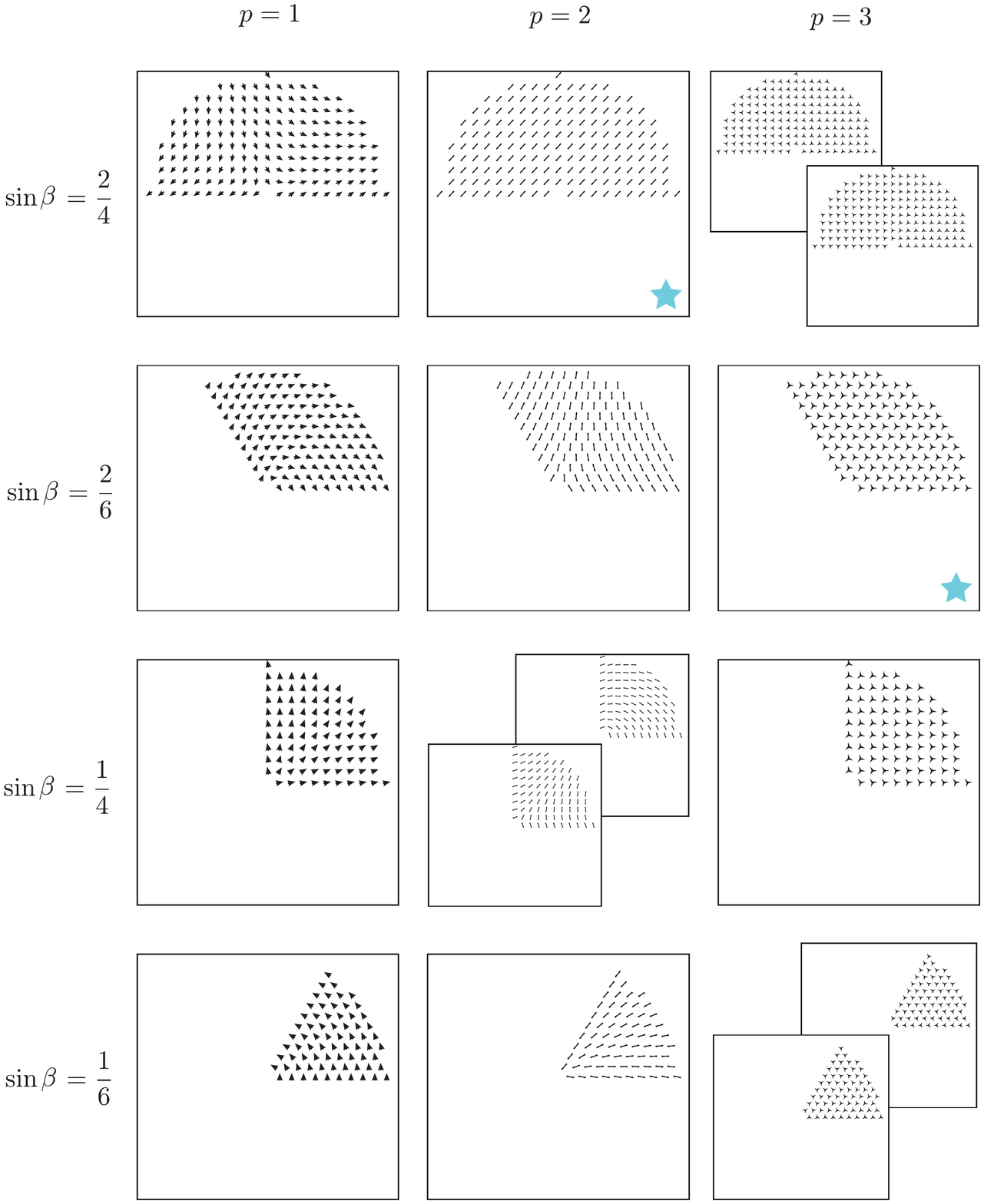}
    % \caption{Caption}
    % \label{fig:my_label}
\end{figure*}

\begin{figure*}
    \centering
    \includegraphics[width=1\textwidth]{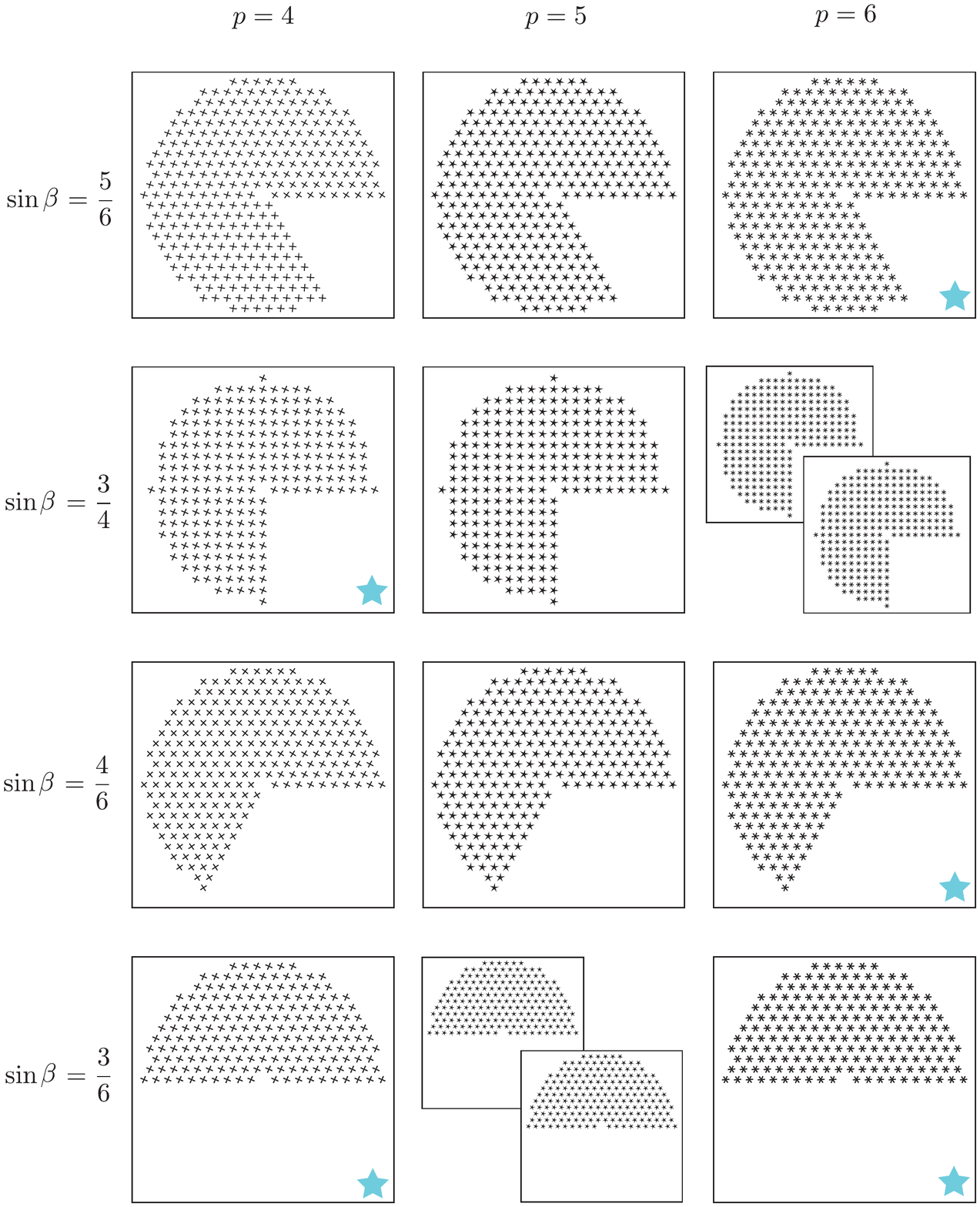}
    % \caption{Caption}
    % \label{fig:my_label}
\end{figure*}

\begin{figure*}
    \centering
    \includegraphics[width=1\textwidth]{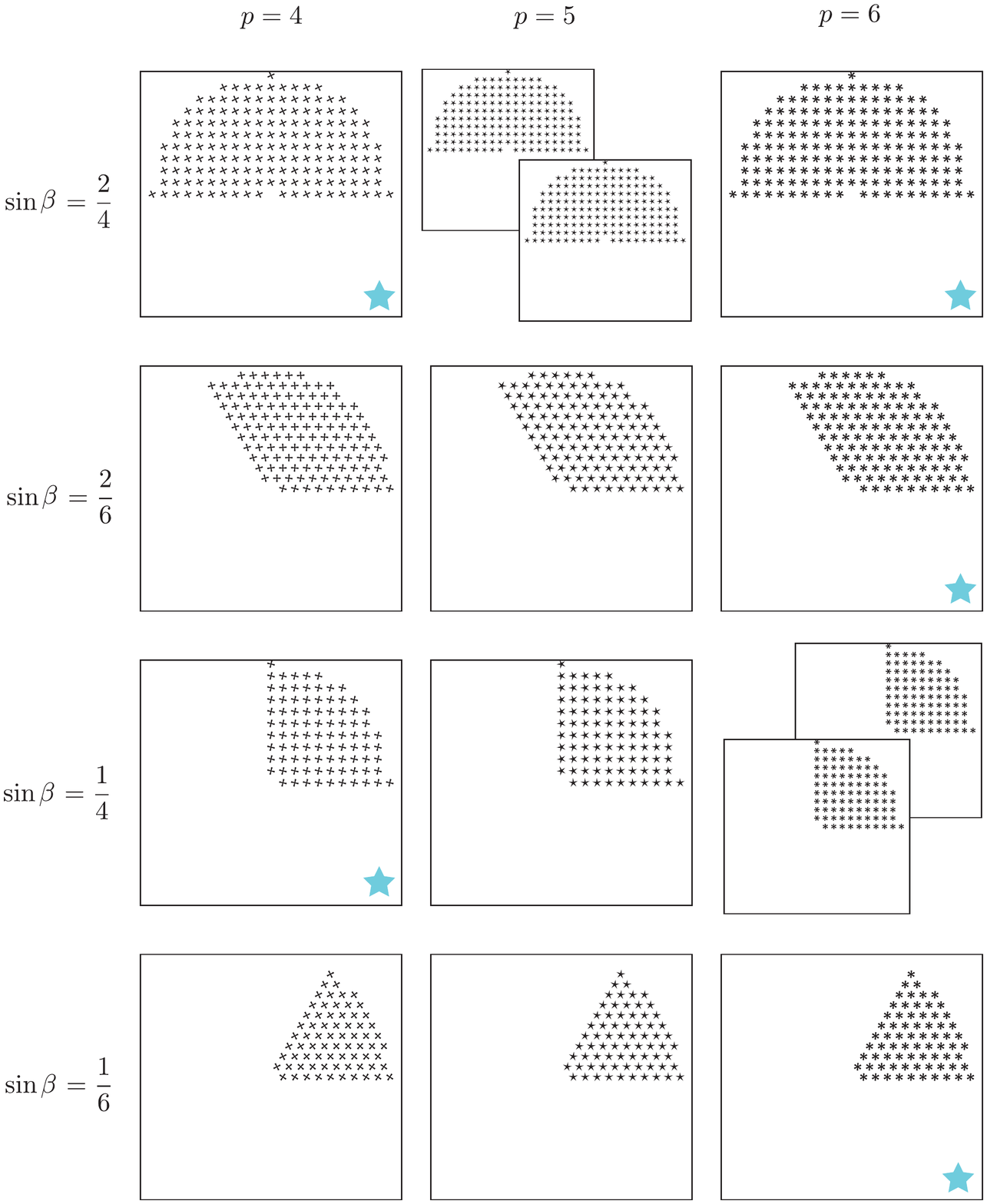}
    % \caption{Caption}
    % \label{fig:my_label}
\end{figure*}

\clearpage

\bibliography{references.bib}
\clearpage

\end{document}